%% file: main.tex
\documentclass{article}
\usepackage{arxiv}
\usepackage{authblk}
\usepackage{amsmath, amssymb, amsthm}
\usepackage{hyperref}
\newtheorem{theorem}{Theorem}
\newtheorem{definition}{Definition}
\newtheorem{lemma}{Lemma}
\newtheorem{corollary}{Corollary}  
\newtheorem{example}{Example}

\newtheorem{proposition}{Proposition}
\input{header/packages}
\input{header/commands}

\title{Functional Reactive Programming With Effects,\\ A More Permissive Approach}
\date{{\tt{frederic.dabrowski@univ-orleans.fr}} \qquad {\tt{jordan.ischard@univ-orleans.fr}}}
\author[1]{Frédéric Dabrowski}
\author[1]{Jordan Ischard}
\affil[1]{Univ.Orléans, INSA Centre Val de Loire, LIFO EA 4022, Orléans, France}
\begin{document}
\maketitle

\begin{abstract}
We introduce a functional reactive programming language that extends \Wormholes{}, an enhancement of \Yampa{} with support for effects. Our proposal relaxes the constraint in \Wormholes{} that restricts all resources to single-use.
Resources are categorized into two kinds: input/output resources and internal resources.
Input/output resources model interactions with the environment and follow constraints similar to those in \Wormholes{}. Internal resources, on the other hand, enable communication between program components and can be used multiple times.
We demonstrate that programs written in our language can be translated into equivalent effect-free \Yampa{} programs, ensuring that our approach remains compatible with existing functional reactive paradigms.
\end{abstract}

\section{Introduction}
\label{sec:introduction}
\input{pages/introduction}

\section{Overview}
\label{sec:overview}
\subsection{\Yampa}
\label{sec:yampa}
\input{pages/yampaoverview}

\subsection{\Wormholes}
\label{sec:wormhole}
\input{pages/wormholeoverview}

\section{Semantic Domains}
\label{sec:preliminary}
\input{pages/preliminary}

\section{\Yampacore}
\label{sec:yampasemantics}
\input{pages/yampasemantics}

\section{The \ourlanguage language}
\label{sec:frp}
\input{pages/frp}

\section{Equivalence}
\label{sec:equivalence}
\input{pages/equivalence}

\section{Conclusion}
\label{sec:conclusion}
\input{pages/conclusion}

\bibliography{references}

\end{document}

%% file: header/packages.tex
\usepackage{cancel}
\usepackage{todonotes}                                 
\usepackage{proof}
\usepackage{xspace}
\usepackage{tikz-cd}
\usetikzlibrary{snakes,arrows.meta,shapes.arrows}
\usepackage[inline]{enumitem}
\usepackage{stmaryrd}
\usepackage{cleveref}
\usepackage{bold-extra}

%% file: header/commands.tex
\def\ourlanguage{{\sc{Molholes}}\xspace}
\def\Yampacore{{\sc{YampaCore}}\xspace}
\def\Yampa{{\sc{Yampa}}\xspace}
\def\Coq{{\sc{Coq}}\xspace}
\def\OCaml{{\sc{OCaml}}\xspace}
\def\FRAN{{\sc{FRAN}}\xspace}
\def\Dunai{{\sc{Dunai}}\xspace}
\def\FRPBearRiver{{\sc{FRPBearRiver}}\xspace}
\def\Haskell{{\sc{Haskell}}\xspace}
\def\Wormholes{{\sc{Wormholes}}\xspace}
\def\FRPNow{{\sc{FRPNow}}\xspace}
\def\Esterel{{\sc{Esterel}}\xspace}
\def\Lustre{{\sc{Lustre}}\xspace}
\def\LucidSynchrone{{\sc{LucidSynchrone}}\xspace}
\def\Signal{{\sc{Signal}}\xspace}
\def\ReactiveML{{\sc{ReactiveML}}\xspace}
\def\SL{{\sc{SL}}\xspace}
\def\ReactiveC{{\sc{ReactiveC}}\xspace}
\newcommand{\letin}[3]{{\tt{let}}~#1 = #2~{\tt{in}}~#3}

\def\mbind{{\mathit{bind}}}
\def\mreturn{{\mathit{return}}}

\newcommand{\funsymbol}[1]{{\texttt{#1}}}

\def\void{\funsymbol{tt}}
\def\true{\funsymbol{true}}
\def\false{\funsymbol{false}}
\def\id{\funsymbol{id}}

\def\mread{\funsymbol{read}}
\def\mwrite{\funsymbol{write}}
\def\readable{\funsymbol{readable}}
\def\writable{\funsymbol{writable}}

\def\aread{\funsymbol{read}^{\dagger}}
\def\areadable{\funsymbol{readable}^{\dagger}}
\def\awrite{\funsymbol{write}^{\dagger}}
\def\awritable{\funsymbol{writable}^{\dagger}}

\def\Stateret{\funsymbol{return}}
\def\Statebind{\funsymbol{bind}}
\def\Stateget{\funsymbol{get}}
\def\Stateset{\funsymbol{set}}

\def\domain{\funsymbol{dom}}

\def\aeval{{\funsymbol{step}^{\dagger}}}
\def\eval{\funsymbol{step}}
\def\run{\funsymbol{run}}

\newcommand{\init}{\funsymbol{init}}
\newcommand{\ainit}{{\funsymbol{init}^{\dagger}}}
\newcommand{\pull}{\funsymbol{pull}}
\newcommand{\push}{\funsymbol{push}}

\newcommand{\typesymbol}[1]{{\mathit{#1}}}

\def\TypeType{\typesymbol{Type}}
\def\VoidType{\typesymbol{Unit}}
\def\BoolType{\typesymbol{Bool}}
\def\NatType{\typesymbol{Nat}}
\def\StreamType{\typesymbol{Stream}}
\def\ListType{\typesymbol{List}}

\def\SfType{\typesymbol{SF}}
\def\sfType{\typesymbol{sf}}

\def\State{\typesymbol{St}}

\def\RSfType{{\typesymbol{RSF}}}
\def\rsfType{{\typesymbol{rsf}}}

\def\RefType{\typesymbol{Ref}}

\def\TagType{\typesymbol{Tag}}
\def\StatusType{\typesymbol{Status}}

\def\AMemoryType{\typesymbol{Memory}^{\dagger}}

\def\CellType{\typesymbol{Cell}}
\def\MemoryType{\typesymbol{Memory}}

\def\ValType{\typesymbol{Val}_{\TypeType}}
\def\ProgType{\typesymbol{Prog}}

\newcommand{\constructorsymbol}[1]{{\texttt{#1}}}

\def\Cons{\constructorsymbol{Cons}}

\def\Arr{\constructorsymbol{Arr}}
\def\Cell{\constructorsymbol{Cell}}
\def\Comp{\constructorsymbol{Comp}}
\def\First{\constructorsymbol{First}}

\def\Loop{\constructorsymbol{Loop}}
\def\Refcons{\constructorsymbol{Ref}}
\def\Get{\constructorsymbol{Get}}
\def\Set{\constructorsymbol{Set}}

\def\sfval{{\mathit{sf}}}
\newcommand{\sgval}{{\mathit{sg}}}

\def\rsfval{\mathit{rsf}}
\newcommand{\sfvaln}[1]{\sfval_{\!#1}}
\newcommand{\rsfvaln}[1]{\rsfval_{\!#1}}
\newcommand{\sgvaln}[1]{\mathit{sg}_{#1}}

\def\Input{\constructorsymbol{Input}}
\def\Internal{\constructorsymbol{Internal}}
\def\Output{\constructorsymbol{Output}}

\def\arr{{\mathit{arr}}}
\def\comp{{\mathit{comp}}}
\def\compop{\mathbin{>\!>\!>}}
\def\first{{\mathit{first}}}
\def\sloop{{\mathit{loop}}}
\def\rsf{{\mathit{rsf}}}
\def\wormhole{\mathit{wormhole}}
\def\get{{\mathit{get}}}
\def\set{{\mathit{set}}}

\def\assoc{\mathit{assoc}}
\def\swap{\mathit{swap}}
\def\unassoc{\mathit{unassoc}}
\def\unassoc{\mathit{unassoc}}
\def\dup{\mathit{dup}}
\def\sdup{\mathit{sdup}}
\def\perm{\mathit{perm}}
\def\step{\mathit{step}}
\def\refget{\mathit{\mathit{refs}_{\get}}}
\def\refset{\mathit{\mathit{refs}_{\set}}}

\def\stackfirst{\mathit{stack}}

\def\complg{\mathit{seq}^{\tt G}_{l}}
\def\comprg{\mathit{seq}^{\tt G}_{r}}
\def\compls{\mathit{seq}^{\tt S}_{l}}
\def\comprs{\mathit{seq}^{\tt S}_{r}}

\def\permlf{\perm_{\mathit{l}}}
\def\permrf{\perm_{\mathit{r}}}

\def\fst{\mathit{fst}}

\newcommand{\category}[1]{{\mathcal{#1}}}
\newcommand{\morphism}[3]{\category{#1}\,#2\,#3}

\def\chost{{\mathcal{C}_{\text{hst}}}}
\def\csf{{\mathcal{C}_{\sfType}}}
\def\crsf{\mathcal{C}_{\rsfType}}

\tikzset{
  every node/.style={
    scale = 0.7
  },
  get/.style={
    dashed,
    single arrow,
    single arrow head extend=0,
    single arrow tip angle=140,
    minimum width = 40pt,
    minimum height = 37pt,
    draw=black,
    fill=white,
    text=black!80,
    shape border uses incircle,
    shape border rotate=-90,
  },
  set/.style={
    dashed,
    single arrow,
    single arrow head extend=0,
    single arrow tip angle=140,
    minimum width = 40pt,
    minimum height = 37pt,
    draw=black,
    fill=white,
    text=black!80,
    shape border uses incircle,
    shape border rotate=90,
  },
  arr/.style={
    minimum width = 40pt,
    minimum height = 30pt,
    draw=black,
    fill=white,
    text=black!80,
  },
  first/.style={
    minimum width = 80pt,
    minimum height = 55pt,
    draw=black,
    fill=gray!20,
    text=black!80,
    rounded corners,
  },
  loopPre/.style={
    minimum width = 80pt,
    minimum height = 55pt,
    draw=black,
    fill=gray!20,
    text=black!80,
  },
  rsf/.style={
    minimum width = 40pt,
    minimum height = 30pt,
    draw=black,dashed,
    fill=white,
    text=black!80
  },
  dfarrow/.style={
    -{Triangle[angle=45:4pt]},
    rounded corners,
  }
}

%% file: pages/introduction.tex
Reactive systems, which maintain continuous interaction with their environment, are essential for modern applications such as user interfaces, simulations, and control systems, as highlighted by Harel and Pnueli who first characterized these systems~\cite{Harel1985}. 
One particularly influential line of research has led to several Haskell libraries, originating from \FRAN (Functional Reactive ANimation), as proposed by Elliott and Hudak~\cite{EH1997:ICFP}. These libraries are often categorized under Functional Reactive Programming (FRP), though the term FRP can also encompass a broader family of languages. They provide a declarative approach to programming reactive systems, where computations are expressed as functions over time-varying values, known as signals or streams.

\FRAN was designed to simplify the development of animation and simulation applications in Haskell. Over time, various proposals have optimized the execution model~\cite{ELL2009:ICFP}, enriched the language~\cite{XK2002:PEPM, WCH2012:HASKELL, SKS2020:PPDP}, and addressed limitations~\cite{BAH2022:JFP, PC2015:ICFP, HCN2002:AFP}. One of FRAN’s most notable drawbacks was its inherent space leak, caused by the system retaining past signal values unnecessarily.
The \FRPNow library, designed by Ploeg and Claessen~\cite{PC2015:ICFP}, 
addresses this issue by imposing restrictions on the use of signals,
allowing the system to discard past values.
The \Yampa library, designed by Hudak {\emph{et al.}}~\cite{HCN2002:AFP} takes a more radical approach by not treating signals as first-class citizens. 
In this approach, programs are arrows, a generalization of monads
introduced by Hughes~\cite{HUG2000:SCP}. Intuitively, arrows are to morphisms what monads are to objects.
They denote stateful transformers, that generate an output and an updated transformer from each input.
The latter is then applied to the next input, continuing the
process iteratively. The arrow structure of programs allows them to be composed to build more complex systems.
This approach aligns \Yampa with synchronous languages such as \Esterel~\cite{BER2000:BOOK}, \Signal~\cite{BLJ1991:SCP}, \Lustre ~\cite{HCR1991:ProcIEEE} and \LucidSynchrone~\cite{lucy:manual06}, which operate based on a logical notion of time. At each step (or tick of the logical clock), the system produces an output depending on the current input, thus capturing the essence of length-preserving synchronous functions as defined by Caspi and Pouzet~\cite{CAP1998:CMCS}.
The declarative nature of \Yampa aligns it more closely with data-flow-oriented programming languages such as
\Signal and \Lustre, as opposed to control-flow-oriented programming languages such as \Esterel.
Libraries for general-purpose programming languages have also been proposed for the latter paradigm, including \ReactiveC~\cite{BOU1991:SPE}, \SL~\cite{BS1996:TSE}, and \ReactiveML~\cite{MP2005:PPDP}.

Despite \Yampa{}’s advantages, managing I/O operations remains challenging due to the lack of support for monadic programming in the language. An early attempt to address this limitation was proposed by Winograd-Cort and Hudak~\cite{WLH2012:PADL, WCH2012:HASKELL}, who introduced resource signal functions (RSFs). RSFs facilitate imperative-style resource access within a functional framework.
Resources can be either global, representing inputs and outputs, or local, representing local resources.
In a pure functional setting, this approach has since been superseded by Monadic Stream Functions (MSFs), introduced by Perez {\emph{et al.}}~\cite{PBN2016:HASKELL}. Roughly speaking, MSFs extend the Arrow framework of \Yampa{} by embedding a monadic structure into the type of signal functions. This concept has been implemented in the \Dunai{} and \FRPBearRiver{} libraries, the latter being a refactored version of \Yampa{}, built on top of \Dunai{}.
Despite these advancements, we believe that \Wormholes{} provides a solid foundation for adapting the model to an impure language such as OCaml, where monads are less naturally integrated. In this context, resources could be represented using mutable references. This originally served as our primary motivation for the work presented in this paper.

\Wormholes provides a simple and elegant way to manage resources. However, the original model imposes strict access constraints, allowing resources to be accessed only once.
For inputs and outputs, this restriction aligns well with the synchronous hypothesis, which states that each computational step should appear as an atomic operation to the environment. In synchronous language terminology, such computations are said to take zero time. However, for local resources, this restriction may be overly rigid.

In this paper, we introduce \ourlanguage, a language that extends the \Wormholes approach by relaxing resource usage constraints. \ourlanguage adopts a three-resource paradigm consisting of inputs, outputs, and internal resources. It drops the single-access restriction on local resources, allowing multiple reads and writes. However, input resources must be read exactly once, and output resources must be written exactly once. The latter property ensures that programs are productive.
We formalize the semantics of \ourlanguage and present a trivial static analysis that ensures correct resource usage. We also establish that well-typed programs can be transformed into equivalent programs with two key properties.
The first property states that all resources, including internals, resources are read and written at most once. The second property states that all read operations occur at the beginning of computation, while all write operations occur at the end. 
This aligns with the synchronous hypothesis.
Finally, we demonstrate that this transformation enables converting our programs into resource-free Yampa programs while preserving their semantics. To achieve this, we formalize \Yampa{}'s semantics and establish the correctness of our transformation using equational reasoning and bisimulation techniques. As an introductory example, we demonstrate 
how bisimulation can be used to establish two well-known properties:
that \Yampa{}'s semantics domain indeed form an Arrow and that \Yampa{} programs admit a kind of normal form.
Notably, the form of programs generated by the first transformation can also be considered a normal forms in
\ourlanguage{}, while the second transformation maps this normal form to a \Yampa program in normal form.
Except for resource accesses, which require specific equalities, the transformations rely on standard equational theories of categories, functors, monads, and arrows.

In \cref{sec:overview}, we provide a brief overview of \Yampa{} and \Wormholes{}.
In \cref{sec:preliminary}, we introduce some basic equational theory of categories, functors, monads, and arrows.
We also recall the definition of coalgebras and bisimulation.
In \cref{sec:yampasemantics}, we formalize a core subset of \Yampa and employ bisimulation techniques to demonstrate that the semantics satisfies Arrows equations and that programs admit a normal form.
In \cref{sec:frp}, we introduce \ourlanguage, formalize its semantics, and prove the correctness of its type system.
In \cref{sec:equivalence}, we present the two transformations mentioned above. 
Finally, in \cref{sec:conclusion}, we summarize our contributions and outline directions for future work.

%% file: pages/yampaOverview.tex
The \Yampa library offers a domain-specific language, built on top of \Haskell, designed
for programming reactive systems. It offers a rich set of transformers to define functions which transform input streams into output streams. These transformers, commonly referred to as signal functions, are represented as instances of the {\tt{Arrow}} type class.
In \Yampa, signal functions have the type \(\sfType\, A\, B\), where \(A\) and \(B\) are \Haskell types representing the types of values carried by input and output signals, respectively.
The primary signal functions include \(\arr\), \(\first\), \(\comp\), and \(\sloop\), with
\(\comp\) typically written in infix notation as \(\cdot \compop \cdot \). 
Their behaviors are outlined below, where \(f\) denotes a \Haskell function of type
\(A \rightarrow B\) and \(\sfval\), \(\sfvaln{1}\) and \(\sfvaln{2}\) represent signal functions. A graphical representation of these signal functions is provided in \cref{figure:example:yampa:arrows} and a formal semantics will be presented in \cref{sec:yampasemantics}.
\input{figures/yampa/example/arrows_overview} 
\begin{itemize}
    \item \(\arr\) has the type \((A \rightarrow B) \rightarrow \sfval\, A\,B\). It
    transforms a \Haskell function into a signal function.
    For example, \(\arr\, (*2)\) represents a signal function that doubles a numeric input. 
    \item \(\first\) has the type 
    \(\sfType\, A\, B \rightarrow \sfType\, (A \times C)\, (B \times C)\).
    Given a signal function \(\sfval\), \(\first\, \sfval\)
    processes the first component of the input signal with \(\sfval\)  while
    leaving the second component unchanged.
    \item \(\comp\) has the type \(\sfType\, A\, B \times \sfType\, B\, C \rightarrow \sfType\, A\, C\). It composes two signal functions sequentially.
    \item \(\sloop\) has the type 
    \(C \rightarrow \sfType\, (A \times C)\, (B \times C) \rightarrow  \sfType\, A\, B\).
    Given a value \(v\) and a signal function \(\sfval\), \(\sloop\, v\, \sfval\)
    is a signal function with internal feedback, enabling stateful computations.
\end{itemize}
    The value \(v\) in \(\sloop\, v\, \sfval\) represents the initial state of the computation.
    In \cref{figure:example:yampa:arrows}, the value $v_i$ represents 
    the dynamically evolving internal state.  
    For example \({\mathit{delay}}\,x = \sloop\, x\, (\arr\,{\swap})\), where 
    \(\swap = \lambda (x,y).(y,x)\), represents a signal function 
    that takes as input a signal carrying values of type \(A\) and produces an output signal with the same values,
    delayed by one step and starting with the initial value \(x\).
    \begin{center}
        \begin{tikzpicture}[scale=0.7]
            \node (start) at (-3,0) {};s
            \node (end) at (3,0) {};
            \node[loopPre, minimum width = 100pt] (loop) at (0,-0.2) {};
            \node[arr] (swap) at (0,0) {$\swap$};
            \draw[dfarrow] (start) to[edge label={\small $v$}, pos=0.3] (swap);
            \draw[dfarrow] (swap) to[edge label={\small $x$}, pos=0.7] (end);
            \draw (swap) to[edge label={\small $(x,v)$}, pos=0.2] (end);
            \draw (start) to[edge label={\small $(v,x)$}, pos=0.8] (swap);
            \draw[rounded corners] (swap) -- (1.25,0) to[edge label={\small $v$}, pos=0.5] (1.25,-1) -- (-1.25,-1) to[edge label={\small $x$}, pos=0.5] (-1.25,0) -- (swap);        
          \end{tikzpicture}  
    \end{center}
A signal function of type \(\sfType\, A\, B\) takes an input of type \(A\) and produces both
an output of type \(B\) and an updated version of itself.
When processing an input stream, the signal function is applied to the stream's head
to produce the current output. The update signal function is then applied to the tail of the stream, allowing the state to evolve step-by-step.

%% file: figures/yampa/example/arrows_overview.tex
\begin{figure}[ht]
  \centering
      
  \begin{tikzpicture}[scale=0.7]

    \node (arrIS) at (-2.25,2) {};
    \node[arr] (arr) at (-0,2) {$\arr\,f$};
    \node (arrOS) at (2.25,2) {};
    \node (arrLegend) at (0,1.1) {$\arr\,f$};

    \draw[dfarrow] (arrIS) to[edge label={$v$}, pos=0.5] (arr);
    \draw[dfarrow] (arr)   to[edge label={$f\,v$}, pos=0.5] (arrOS);

    \node (firstIS) at (-2.8,-1) {};
    \node[first] (first) at (0,-1.2) {};
    \node[arr] (firstArr) at (0,-1) {$\sfval$};
    \node (firstOS) at (2.95,-1) {};
    \node (firstLegend) at (0,-2.5) {$\first\,\sfval$};

    \draw[dfarrow] (firstIS) to[edge label={$(v_1,v_2)$}, pos=0.45] (-1.25,-1) to (-1.25,-2) to[edge label={$v_2$}, pos=0.5] (1.25,-2) to (1.25,-1) to[edge label={$(v_1',v_2)$}, pos=0.5] (firstOS);
    \draw[dfarrow] (firstIS) to[edge label={$v_1$}, pos=0.85] (firstArr);
    \draw[dfarrow] (firstArr)   to[edge label={$v_1'$}, pos=0.15] (firstOS);

    \node (compIS) at (6.25,2) {};
    \node[arr] (arr1) at (8,2) {$\sfvaln{1}$};
    \node[arr] (arr2) at (10.5,2) {$\sfvaln{2}$};
    \node (compOS) at (12.25,2) {};
    \node (compLegend) at (9.25,1.1) {$\sfvaln{1} \compop \sfvaln{2}$};

    \draw[dfarrow] (compIS) to[edge label={$v$}, pos=0.5] (arr1);
    \draw[dfarrow] (arr1) to[edge label={$v'$}, pos=0.5] (arr2);
    \draw[dfarrow] (arr2) to[edge label={$v''$}, pos=0.5] (compOS);

    \node (loopIS) at (6.25,-1) {};
    \node[loopPre,minimum width = 120pt] (loop) at (9.25,-1.2) {};
    \node[arr] (loopArr) at (9.25,-1) {$\sfval$};
    \node (loopOS) at (12.25,-1) {};
    \node (loopLegend) at (9.25,-2.5) {$\sloop\,v_0\,\sfval$};

    \draw[dfarrow] (loopIS) to[edge label={$v$}, pos=0.2] (loopArr);
    \draw[dfarrow] (loopArr)   to[edge label={$v'$}, pos=0.8] (loopOS);
    \draw[rounded corners] (loopArr)  to (10.5,-1) to[edge label={\small $v_{i+1}$}, pos=0.5] (10.5,-2) to (8,-2) to[edge label={\small $v_i$}, pos=0.5] (8,-1) to (loopArr);

  \end{tikzpicture}
  
  \caption{Graphical representation of primary signal functions}
  \label{figure:example:yampa:arrows}
\end{figure}
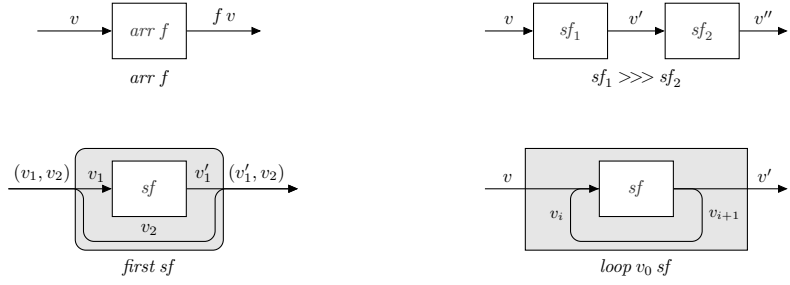

%% file: pages/wormholeoverview.tex
The \Wormholes library extends \Yampa by introducing new constructs for controlled side effects.
At the core of this extension is the concept of resources, which are identifiers attached to 
side effects.
Each resource can be accessed at most once.
Signal functions, or resource signal functions, have types of the form 
\(A \overset{R}{\rightsquigarrow} B\), 
where \(A\) and \(B\) are the types of input and output values, respectively.
The annotation \(R\) denotes the set of resources involved in the computation and 
is used by the type system to ensure correct resource usage.
To access resources, \Wormholes introduces a construct, named $\rsf$, which creates
a signal function $\rsf[r]$ from a resource identifier \(r\).
This signal function takes an input value to be consumed by the environment and returns an output value
produced by the environment. A resource \(r\) has a type \(\langle A, B \rangle\), 
where \(A\) and \(B\) specify the types of inputs and outputs, respectively.
Given such a resource \(r\), the signal function \(\rsf[r]\) has type \(A \overset{\{r\}}{\rightsquigarrow} B\). 

Resources are classified as either unbound or bound. 
Unbound resources represent interactions with the
real world, while bound resources represent communication channels between components of the program.
To manage bound resources, \Wormholes introduces the construct 
\[
\wormhole[r_{get};r_{set}](t_0;t)\] which 
binds the resources \(r_{get}\) and \(r_{set}\) within the term \(t\). 
The term $t_0$ represents the initial value of the resource
\(r_{get}\) while \(r_{set}\) is initialized with ${\tt{tt}}$.
The first resource is used for reading a value from the channel and must have a type of the
form \(\langle \VoidType, A \rangle\), while the second resource is used for writing a value to the channel 
an must have a type of the form \(\langle A, \VoidType \rangle\).
The type of such a term matches that of \(t\), except that 
\(r_{get}\) and \(r_{set}\) are removed.
Regardless of whether resources are bound or unbound, they are single-use, meaning they can be accessed at most once.

In addition to the $\rsf$ and $\wormhole$ constructs, \Wormholes provides
the signal functions $\arr$, $\first$ and $\comp$ which are similar to their
\Yampa counterparts. 
Unlike \Yampa{}, \Wormholes does not include a $\sloop$ function.
However, as demonstrated by Winograd-Cort and Hudak, it can be emulated. Specifically, a term of the form 
$\sloop\,v_0\,t$ can be encoded as
\begin{align*}
  \wormhole[r_{get},r_{set}](v_0;&
  \,\arr\,(\lambda x.(\void,x)) \compop \first\,(\rsf[r_{get}]) \compop\\ 
  &\arr\,(\lambda(x,y).(y,x)) \compop t \compop \arr\,(\lambda(x,y).(y,x)) \compop\\
  &\first\,(\rsf[r_{set}]) \compop \arr\,{\mathit{snd}})
\end{align*}
where $r_{get}$ and $r_{set}$ are fresh resources, $v_0$ is the initial value of the loop, 
and $t$ is its body.
\input{figures/yampa/example/arrows_comparison}

The use of resources in \Wormholes enhances modularity. Consider
a program that performs two tasks sequentially:  the first task sends a value to the environment, while the second task receives a value from the environment.
In \Yampa, both the input stream and the output stream must carry a pair, as illustrated in the top schema 
of \cref{figure:example:arrow:comparison}.
In contrast, an equivalent \Wormholes program requires only the first component, while
defining a resource $r$ to represent the interaction "I provide my first result in exchange for the expression needed for my second task",
as depicted in \cref{figure:example:arrow:comparison}.

To enforce an affine use of resources at the semantics level, 
\Wormholes introduces a memory cell for each resource, see
\cref{figure:example:arrow:rsf} where the resource \(r\) is
associated with the memory cell \(c\).
Each memory cell carries a tag that indicates whether its associated resource has been used. 
The cell consists of tow sub-cells, one for the input and one for the 
output. However, a cell cannot hold both pieces of information simultaneously. 
The tag is inferred from the shape of the cell.
If the left sub-cell contains an element while the right 
sub-cell contains ($\texttt{-}$), the cell is accessible.
At the beginning of each computation step, all cells are accessible. 
When a resource is accesses through the $\rsf$ construct, 
the state of cell updates accordingly. 
The type system of \Wormholes ensures correct resource usage by tracking their accesses.
\input{figures/yampa/example/arrows_rsf}

The problem outlined in this section naturally boils down to the usual challenges of handling effects in a purely functional language. Monads provide an effective solution, and as mentioned in the introduction, \Yampa{}'s model has been adapted to support them. As we also noted, it is with the objective of designing a library for an impure functional language that we build upon the \Wormholes{}'s model.

%% file: figures/yampa/example/arrows_comparison.tex
\begin{figure}[ht]
  \centering
      
  \begin{tikzpicture}[scale=0.7]

    \node (start) at (-2.5,0) {};
    \node (end) at (14.6,0) {};

    \node[first, minimum width = 185pt, minimum height = 55pt] (first) at (2,-0.3) {};
    \node[first, minimum width = 185pt, minimum height = 55pt] (first) at (9.8,0.3) {};
    \node[arr] (A) at (0.1,0) {$t_1$};
    \node[arr] (B) at (2.5,0) {$t_2$};
    \node[] (CD) at (4.4,0) {$\ldots$};
    \node[arr] (E) at (7.9,0) {$t_3$};
    \node[arr] (F) at (10.3,0) {$t_4$};
    \node[] (GHI) at (12.2,0) {$\ldots$};

    \draw[dfarrow] (start) to[edge label={\small $(v_1,v_2)$}, pos=0.3] (A);
    \draw[dfarrow] (A) to (B);
    \draw[dfarrow] (B) to (CD);
    \draw[draw=white] (CD) to[edge label={\small $v_2$}, pos=0.91]  (E);
    \draw (CD) to[edge label={\small $v'_1$}, pos=0.1]  (E);
    \draw (GHI) to[edge label={\small $v'_2$}, pos=0.05]  (end);
    \draw[dfarrow] (E) to (F);
    \draw[dfarrow] (F) to (GHI);
    \draw[dfarrow] (start) to[edge label={\small $v_1$}, pos=1.05] (-1.1,0) 
                           to (-1.1,-1) 
                           to[edge label={\small $v_2$}, pos=0.5] (5.1,-1) 
                           to (5.1,0) 
                           to[edge label={\small $(v'_1,v_2)$}, pos=0.5] (6.7,0)
                           to (6.7,0.8)
                           to[edge label={\small $v'_1$}, pos=0.5] (12.9,0.8) to (12.9,0)
                           to[edge label={\small $(v'_1,v'_2)$}, pos=0.5] (end)
                           ;

    \node (start1) at (-2.5,-4) {};
    \node (end1) at (14.6,-4) {};

    \node[arr] (A1) at (-0.5,-4) {$t_1$};
    \node[arr] (B1) at (2,-4) {$t_2$};
    \node[] (CD1) at (4,-4) {$\ldots$};
    \node[rsf] (R) at (6,-4) {$\rsf[r]$};
    \node[arr] (E1) at (8.5,-4) {$t_3$};
    \node[arr] (F1) at (11,-4) {$t_4$};
    \node[] (GHI1) at (13,-4) {$\ldots$};
    \node[] (dt) at (6,-2.3) {$\ldots$};

    \draw[dfarrow] (start1) to[edge label={\small $v_1$}, pos=0.6] (A1);
    \draw[dfarrow] (A1) to (B1);
    \draw[dfarrow] (B1) to (CD1);
    \draw[dfarrow,dashed] (5.7,-3.48) to[edge label={\small $v'_1$}, pos=0.5] (5.7,-2.5);
    \draw[dfarrow,dashed] (6.3,-2.5) to[edge label={\small $v_2$}, pos=0.53] (6.3,-3.48);
    \draw[dfarrow] (CD1) to[edge label={\small $v'_1$}, pos=0.5] (R);
    \draw[dfarrow] (R) to[edge label={\small $v_2$}, pos=0.4] (E1);
    \draw[dfarrow] (E1) to (F1);
    \draw[dfarrow] (F1) to (GHI1);
    \draw[dfarrow] (GHI1) to[edge label={\small $v'_2$}, pos=0.4]  (end1);
      
  \end{tikzpicture}
  
  \caption{Equivalent programs in \Yampa (top) and in \Wormholes (bottom)}
  \label{figure:example:arrow:comparison}
\end{figure}
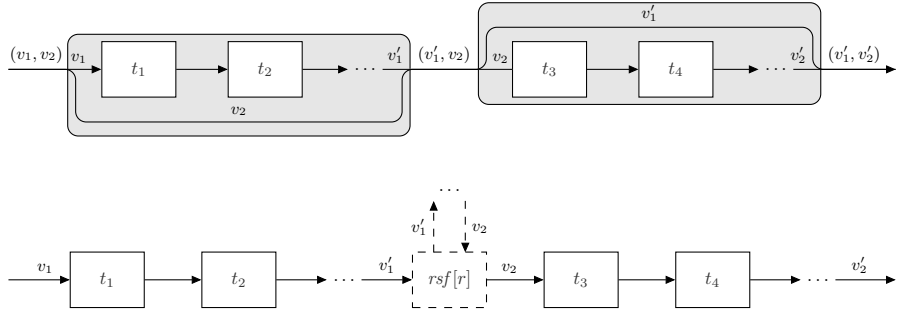

%% file: figures/yampa/example/arrows_rsf.tex
\begin{figure}[ht]
  \vspace*{-3mm}

  \centering
      
  \begin{tikzpicture}[scale=0.7]

    \node (start) at (-3,0) {};
    \node (end) at (3,0) {};
    \node (c) at (-2,-2.1) {c};
    \node (c1) at (2,-2.1) {c};
    \node[draw=black,fill=white,minimum width = 30pt,minimum height = 20pt,text=black!80] (cellget) at (-2.5,-1.5) {\true};
    \node[draw=black,fill=white,minimum width = 30pt,minimum height = 20pt,text=black!80] (cellget1) at (-1.5,-1.5) {\texttt{\_}};
    \node[draw=black,fill=white,minimum width = 30pt,minimum height = 20pt,text=black!80] (cellgetb) at (2.5,-1.5) {2};
    \node[draw=black,fill=white,minimum width = 30pt,minimum height = 20pt,text=black!80] (cellget1b) at (1.5,-1.5) {\texttt{\_}};
    \node[rsf] (rsf) at (0,0) {$\rsf[r]$};
    \draw[-{Triangle[angle=45:4pt]},dashed,rounded corners] (cellget1) to (-0.5,-0.55);
    \draw[-{Triangle[angle=45:4pt]},dashed,rounded corners] (0.5,-0.55) to (cellget1b);
    \draw[-{Triangle[angle=45:4pt]},rounded corners] (rsf) to[edge label={\small \true}, pos=0.3] (end);
    \draw[-{Triangle[angle=45:4pt]},rounded corners] (start) to[edge label={\small 2}, pos=0.7] (rsf);
    \draw[-{Triangle[angle=45:4pt]},decorate,decoration={snake,amplitude=.4mm,segment length=2mm,post length=1mm}] (cellget1) to[edge label={\small update state}, pos=0.5] (cellget1b);

  \end{tikzpicture}
  
  \caption{Graphical representation a $\rsf$ term computation.}
  \label{figure:example:arrow:rsf}
\end{figure}
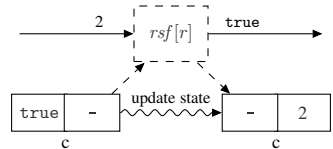

%% file: pages/preliminary.tex
\def\ArrC{\text{Arr}}
\def\arrC{\text{arr}}
\def\firstC{\text{first}}
\def\sloopC{\text{loop}}
In sections \cref{sec:yampasemantics} and \cref{sec:frp}, we formalize the semantics of \Yampa and \ourlanguage, 
and in \cref{sec:equivalence}, we establish the correctness of various program transformations.
These transformations occur both within each language and between the two.
Both languages are built upon a common abstract host language, axiomatized at the semantics level by 
a collection of typed values and typed functions forming a cartesian category \(\chost\).
The cartesian requirement ensures that product types and functions over product types are available in the host language,
which is essential for our transformations.

Programs in \Yampa{} and \ourlanguage{} represent length-preserving synchronous functions. 
Their semantics is defined by iteratively applying a step function.
The semantic domains of these step functions can be modeled as terminal coalgebras of a 
family of functors.
This family, indexed by types of the host language, is defined as follows:
\[
\begin{array}{lcll}
    F_{AB}\,X &=& A \rightarrow (B \times X)\\
    F_{AB}\,\step &=& \lambda a. (b,F_{AB}\,g) & {\text{where }} (b,g) = \step\,a
\end{array}
\]
Given two types \(A\) and \(B\) and a coalgebras \((X,f)\) for \(F_{AB}\),
\(X\) represents a type of transformers and \(f\) is a morphism mapping a transformer to a step function.
Intuitively, this function applies the transformer to an input value of type \(A\), producing an output 
value of type \(B\) along with a new transformer. 
In the case of \ourlanguage{}, the coalgebra incorporates a state monad to express the semantics of effects.
A transformation from \ourlanguage{} to \Yampa{} is then expressed as a coalgebra morphism between their
respective semantics domains. In both cases, the carrier of the coalgebra forms
an arrow, as defined by Hughes~\cite{HUG2000:SCP}. The correctness of the transformation is established by
applying bisimulation proof techniques between coalgebras of \(F_{AB}\), we obtain a powerful method for
establishing behavioral equivalence (bisimilarity) both within a single language and between the two languages.
Intuitively, bisimilarity between two transformers expresses the idea that they exhibit the same behavior.
In the context of \Yampa{} and \ourlanguage{}, this means that, for any given input, the transformers
produce the same output and that their updated transformers remain bisimilar. Bisimilarity is
required because our semantics domain consists of corecursive functions. Proving equivalence between
corecursive functions requires coinductive reasoning techniques, which do not apply to 
Leibniz equality.

In the remainder of this section, we provide a brief overview of the equational theories of categories, 
functors, monads, and arrows. Additionally, we define bisimulation between elements of two coalgebras 
of the functor \(F_{AB}\).
This section, which may be skimmed on a first reading, presents only the essential material needed to
justify the transformations applied later in the the paper.
For an in-depth introduction to these concepts, we refer the reader to the works of Spivak~\cite{Spivak2014},
Jacobs~\cite{Jacobs2016}, and Sangiorgi~\cite{Sangiorgi2011}. Our presentation of monads and arrows follows the
functional programming style, which differs slightly from the category-theoretic approach.
For a comprehensive introduction to monads in the context of functional programming, 
we recommend Wadler's paper~\cite{Wadler1990}, while Hughes's paper~\cite{HUG2000:SCP} introduces arrows in a similar manner. A comprehensive study of Monads can be found in Moggi's seminal paper~\cite{Moggi1991}.
An in-depth study of the relationship between arrows and Freyd categories can be found in Atkey's paper~\cite{Atkey2011}.

\subsection{Categories}
A category \(\category{C}\) consists of a collection of objects, which we denote by \(\category{C}\),
and a collection of morphisms for each pair of objects
\(A\) and \(B\), which we denote by \(\morphism{C}{A}{B}\). Additionally, a category \(\category{C}\)
includes:
\begin{itemize}
    \item a family of identity morphisms \(id_A\) for each object \(A\)
    \item a family of composition operators 
    \(\circ_{ABC} : \morphism{C}{B}{C} \times \morphism{C}{A}{C} \rightarrow \morphism{C}{A}{C}\) 
    for each triple of objects \(A\), \(B\), and \(C\).
\end{itemize}
These morphisms and operators must satisfy the following laws:
\begin{subequations}\label{eq:categorylaws}
    \begin{align}
        f \circ id &= f \label{eq:categorylaw1}\\
        id \circ f &= f\label{eq:categorylaw2}\\
        (f \circ g) \circ h &= f \circ (g \circ h)\label{eq:categorylaw3}
    \end{align}
\end{subequations}
An object \(A\) is terminal if for all objects \(B\), there exists a unique morphism
\(f : \morphism{C}{B}{A}\). Terminal objects are unique up to isomorphism.
A cartesian category is a category with a terminal object and binary products for all pairs of objects.
For a pair of objects \(A\) and \(B\), the product of \(A\) and \(B\) is an object \(A \times B\)
with two projection morphisms \(\pi_1 : \morphism{C}{A \times B}{A}\) and
\(\pi_2 : \morphism{C}{A \times B}{B}\) such that for all objects \(C\) and morphisms
\(f : \morphism{C}{C}{A}\) and \(g : \morphism{C}{C}{B}\), there exists a unique morphism
\(\langle f, g \rangle : \morphism{C}{C}{A \times B}\) such that 
\[\pi_1 \circ \langle f, g \rangle = f\ {\text{ and }} \pi_2 \circ \langle f, g \rangle = g\].
Given a cartesian category, we define the following morphisms which will be used through
the rest of the paper:
\begin{itemize}
    \item \(\dup = \lambda x.(x,x)\) and \(\sdup = \lambda (x,y).((x,y),y)\)
    \item \(\swap = \lambda (x,y).\,(y,x)\)
    \item \(\assoc = \lambda ((x,y),z).\,(x,(y,z))\) and \(unassoc = \lambda (x,(y,z)).\,((x,y),z)\)
    \item \(\perm = \lambda ((x,y),z).\,((x,z),y)\)
    \item \(f\times g = \lambda (x,y).(f\,x, g\,y)\)
\end{itemize}

\subsection{Functors and Monads}
A functor between a category \(\category{C}\) and a category \(\category{D}\)
consists of a mapping of objects
\(F : \category{C} \rightarrow \category{D}\) and a mapping of morphisms
\(F : (A \rightarrow B) \rightarrow F\,A \rightarrow F\,B\) for each pair of objects
\(A\) and \(B\) such that \(F\) preserves identities and compositions.
\begin{subequations}\label{eq:functorlaws}
    \begin{align}
        F\,id &= id\label{eq:functorlaw1}\\
        F\,(g \circ f) &= F\,g \circ F\,f \label{eq:functorlaw2}
    \end{align} 
\end{subequations}
When \(\category{C} = \category{D}\), the functor \(F\) is called an endofunctor.
A monad in a category \(\category{C}\) consists of an endofunctor
\(M : \category{C} \rightarrow \category{C}\) and two indexed family of operators 
\(return_{A} : A \rightarrow M\,A\) and
\(bind_{AB} : M\,A \rightarrow (A \rightarrow M\,B) \rightarrow {M\,B}\) 
for each pair of objects \(A\) and \(B\). These morphisms must satisfy the following laws:
\begin{subequations}\label{eq:monadlaws}
    \begin{align}
        \mbind\,(\mreturn\,a)\,f &= f\,a \label{eq:monadlaw1}\\
        \mbind\,m\,\mreturn &= m \label{eq:monadlaw2}\\
        \mbind\,(\mbind\,f\,g)\,h &= \mbind\,f\,(\lambda x.\mbind\,(g\,x)\,h) \label{eq:monadlaw3}
    \end{align}
\end{subequations}
A monad in a category \(\category{C}\) gives rise to a Kleisli category \(\category{C}_M\) 
with has the same objects as 
\(\category{C}\), but with morphisms defined as
\(\category{C}_M\,A\,B = \category{C}\,A\,(M\,B)\)
for each pair of objects \(A\) and \(B\). 
Identities are given by \(\mreturn\) and compositions are defined by
\(g \circ f = \lambda a.\,\mbind\,(f\,a)\,g\).

\subsection{Arrows}
\label{sec:arrows}
An arrow in a cartesian category \(\category{C}\) consists of a mapping of objects 
\(\ArrC : \category{C} \rightarrow \category{C} \rightarrow \category{C}\) with three indexed families
of operations:
\begin{equation}
    \begin{array}{lcl}
        \arr_{AB} &:& (A \rightarrow B) \rightarrow \ArrC\,A\, B\\
        \compop_{ABC} &:& 
        \ArrC\,A\,B \times \ArrC\,B\,C \rightarrow \ArrC\,A\,C\\
        \first_{ABC} &:& \ArrC\,A\, B \rightarrow \ArrC\,(A \times C)\, (B \times C)\\
    \end{array}
\end{equation}
Arrows are elements of \(\ArrC\,A\,B\) for each pair of objects \(A\) and \(B\).
The operations above must satisfy the laws enumerated in \cref{eq:arrowlaws}, where 
we denote arrows by \(a\), \(b\), and \(c\), and morphisms in the category \(\category{C}\)
by \(f\) and \(g\). 
\begin{subequations}\label{eq:arrowlaws}
    \begin{align}
    \arrC\,id \compop a &= a \label{eq:arrowlaw1}\\
    a \compop \arrC\,id &= a \label{eq:arrowlaw2}\\
    (a \compop b) \compop c &= a \compop (b \compop c)\label{eq:arrowlaw3}\\
    \arrC\,(g \circ f) &= \arrC\,f \compop \arrC\,g\label{eq:arrowlaw4}\\
    \firstC\,a \compop \arrC\,\pi_1 &= \arrC\,\pi_1 \compop a\label{eq:arrowlaw5}\\
    \firstC\,a \compop \arrC\,(id \times f) &= \arrC\,(id \times f) \compop \firstC\,a\label{eq:arrowlaw6}\\
    \firstC (\firstC\,a) \compop \arrC\,{\assoc} &= \arrC\,{\assoc} \compop \firstC\,a\label{eq:arrowlaw7}\\ 
    \firstC\,(\arrC\,f) &= \arrC\,(f \times id)\label{eq:arrowlaw8}\\
    \firstC (a \compop b) &= \firstC\, a \compop \firstC\, b\label{eq:arrowlaw9}
\end{align}
\end{subequations}
Each Arrow \(\ArrC\) on a category \(\category{C}\) gives rise to a category \(\category{C}_{\ArrC}\)
with the same objects as \(\category{C}\) and morphisms defined by
\(\category{C}_{\ArrC}\,A\,B = \ArrC\,A\,B\)
for each pair of objects \(A\) and \(B\).
The identity of an object $A$ is \(\arrC\,{\mathit{id}}_A\) and composition is defined as
\(g \circ f = f \compop g\).
This structure forms a category that follows
\cref{eq:arrowlaw1,eq:arrowlaw2,eq:arrowlaw3}. Additionally, \cref{eq:arrowlaw4} establishes that
\(\arrC\) is an identity-on-objects functor from the initial category to the category of arrows.
An arrow with loops refers to an arrow equipped with a feedback loop which satisfies additional laws.
\begin{subequations}\label{eq:arrowlooplaws}
    \(\sloop_{ABC} : C \rightarrow \ArrC\,(A \times C)\, (B \times C) \rightarrow \ArrC\,A\,B\)
    \begin{align}
        \sloopC\,c\,(\firstC\,a \compop\, b) &= a \compop \sloopC\,c\,b \label{eq:arrowlooplaw1}\\
        \sloopC\,c\,(a \compop \firstC\,b) &= \sloopC\,c\,a \compop b \label{eq:arrowlooplaw2}\\
        \sloopC\,c\,(\sloopC\,d\,a) &= \sloopC\,(c,d)\,(\arrC\,\unassoc \compop a \compop \arrC\,\assoc)
        \label{eq:arrowlooplaw3}
    \end{align}
\end{subequations}
\subsection{Bisimulations}
Given an endofunctor \(F\) of a category \(C\), a coalgebra for \(F\) (or a $F$-algebra) is a pair \((X, f)\) 
where \(X\) and \(f : X \rightarrow F\,X\) are an object and a morphism in 
the category \(\category{C}\), respectively.
A bisimulation is a relation between elements of two coalgebras of a functor.
For our purposes, it is sufficient to restrict the definition to the functor \(F_{A,B}\), 
defined as:
\[
    \begin{array}{lcl}
        F_{A,B}\,X &=& A \rightarrow (B \times X)\\
        F_{A,B} &:& (X \rightarrow Y) \rightarrow (F_{A,B}\,X \rightarrow F_{A,B}\,Y)\\
        F_{A,B}\,f &=& \lambda {\step}\,a. \letin{(b,x)}{{\step}\,a}{(b,f\,x)} 
    \end{array}
\]
Let \({\step}_X : X \rightarrow F_{A,B}\,X \) 
and \({\step}_Y : Y \rightarrow F_{A,B}\,Y \) be two coalgebras.
A bisimulation between \(X\) and \(Y\) is a relation \(R\) such that
for all \(x \in X\) and \(y \in Y\) with \(R \,x\, y\), the following holds:
\begin{center}
whenever
\(\step\,x\ a = (b,x')\), there exists \(y'\) such that 
\(\step\,y\ a = (b,y')\) and \(R\,x'\,y'\)
\end{center}
Bisimilarity, denoted \(x \sim y\), is the largest bisimulation between \(X\) and \(Y\).
Thus, to demonstrate that two elements are bisimilar, it suffices to exhibit a bisimulation
that relates them.
Bisimilarity between \(X\) and \(Y\) enjoys the following properties:
\begin{subequations}\label{eq:bisimlaws}
    \begin{align}
        x \sim x\\
        x \sim y \Rightarrow y \sim x\\
        x \sim y \land y \sim z \Rightarrow x \sim z
    \end{align}
\end{subequations}
In particular, the relation \(\sim\) is an equivalence relation when \(X=Y\).

It is worth noting that this notion of bisimulation aligns more closely with simulation in the terminology of process algebras.
Bisimulation requires that both elements simulates each other.
However, when considering systems described by total functions, the two notions coincide.
This holds true for the semantics of well-typed programs in the formalization of
both \Yampa{} and \ourlanguage{}.

%% file: pages/yampasemantics.tex
In this section, we present a formal semantics for the kernel language \Yampacore{}, which captures the core concepts of \Yampa{}.
This semantics is used in \cref{sec:equivalence}, where we show how \ourlanguage{} programs, 
as defined in \cref{sec:frp}, can be translated into \Yampacore{} programs.
Additionally, we demonstrate how bisimulation proofs can be used to establish that \Yampacore{} effectively defines an arrow with loops. Furthermore, we show that any program can be transformed into an
equivalent normal form. This well-known result follows directly from the equational theory of arrows with loops.

We define \Yampacore{} as a two-level language: the first level consists of a host language whose semantics
is provided by \(\chost\), while the second level is a domain-specific language for stream processing.
Programs represent total functions over streams of values, where streams are modeled as an indexed family of types
\(\StreamType\, A\), with \(A\) being a type in the host language.
Notably, we do not assume that stream types are types of the host language, reflecting the fact that streams are not treated as first-class citizens in \Yampa{}.
Stream types are defined by the following indexed family:
\[
\begin{array}{lcl}
    \StreamType\,A &=& \Cons : A \times \StreamType\,A \rightarrow \StreamType\,A
\end{array}
\]
The type \(\StreamType\, A\) is isomorphic to the type of infinite lists of elements of type \(A\).
It serves as the carrier of a final coalgebra, characterized by \(\lambda\,(\Cons\,a\,l) = (a,l)\),
for the functor \(F_A\,X = A \times X\).
A bisimulation over streams is defined as a relation \(R\) such that for all \(s\) and \(s'\) such that \(R\,s\,s'\):
\begin{center}
    if \(s = \Cons\,a\,s_1\) then there exists \(s_1'\) such that \(s' = \Cons\,a\,s_1'\) and \(R\,s_1\,s_1'\)
\end{center}
Bisimilarity is defined, as usual, as the largest bisimulation.

\subsection{Syntax and Semantics}
We define the syntax of \Yampacore{} using typed terms, which are elements of
algebras \(\SfType\,A\,B\), where \(A\) and \(B\) are types of the host language. 
The definition of \(\SfType\) aligns with the informal description of \Yampa{} provided
in \cref{sec:yampa}.
    \begin{equation*}
        \begin{array}{lcl}
            \SfType &=& \mid \Arr :\forall A~B.~ (A \rightarrow B) \rightarrow \SfType\, A\, B\\
            && \mid \Comp : \forall A~B~C.~\SfType\, A\, B \times \SfType\, B\, C \rightarrow \SfType\, A\, C\\
            && \mid \First : \forall A~B~C.~\SfType\, A\, B \rightarrow \SfType\, (A \times C)\, (B \times C)\\
            && \mid \Loop : \forall A~B~C.~C \times \SfType\, (A \times C)\, (B \times C) \rightarrow \SfType\, A\, B
        \end{array}
    \label{eq:yampa-syntax}
    \end{equation*}
For example, the term \(\Loop\,v\,(\Arr\,{\mathit{swap}})\) represents the signal function
\({\mathit{delay}}\,v\) where \(delay\) refers to the function
defined in \cref{sec:yampa}.
The semantic domain of terms is defined by the coinductive type:
\[\sfType\,A\,B = A \rightarrow (B \times \sfType\,A\,B)\]
which constitutes a terminal coalgebra \((\sfType\,A\,B, id)\) 
for the functor \(F_{A,B}\,X = A \rightarrow (B \times X)\), as introduced
in \cref{sec:preliminary}.

The stepwise semantics of terms is defined by the recursive function 
\(\eval\).
Its definition relies on the corecursive functions \(\arr\), \(\comp\), \(\first\) and \(\sloop\),
as detailed below.
\begin{equation*}
    \begin{array}{lclll}
        \eval &::& \SfType\, A\, B \rightarrow \sfType\,A\,B\\
        \eval\, (\Arr\, f) &=& \arr\,f
        \\
        \eval\, (\Comp\, \sfvaln{1}\,\sfvaln{2}) &=& 
        \comp\, (\eval\, \sfvaln{1})\,(\eval\, \sfvaln{2})\\
        \eval\, (\First\, \sfval)  &=& \first\,(\eval\,\sfval)
        \\
        \eval\, (\Loop\, \sfval\, v)&=& \sloop\, v\, (\eval\, \sfval)
        \\\\
        \arr\, f &=& \lambda x.(f\, x, \arr\, f)\\
        \comp\,\sfvaln{1}\,\sfvaln{2} &=& 
        \lambda a. (c, \comp\,\sfvaln{1}'\, \sfvaln{2}')
        \\
        && {\text{where }} (b,\sfvaln{1}') = \sfvaln{1}\,a {\text{ and }} (c,\sfvaln{2}') = \sfvaln{2}\,b
        \\
         \first\, \sfval &=& \lambda (x,z).((y,z), \first\, \sfval') \text{ where } y, \sfval' = \sfval\, x\\
        \sloop\, v\, \sfval &=& \lambda x.(y, \sloop\, v'\, \sfval') \text{ where } 
        (y,v'),\sfval' = \sfval\, (x,v)
    \end{array}
\end{equation*}
For example, the semantics of the term \(\Loop\,b\,(\Arr\,{\mathit{swap}})\) is given by
\[
    \begin{array}{lcl}
        \eval\,(\Loop\,b\,(\Arr\,{\mathit{swap}})) &=& \sloop\,b\,(\arr\,{\mathit{swap}})\\
        &=&\lambda a. (b,\,\sloop\,a\,(\arr\,{\mathit{swap}}))\\
        &=&\lambda a. (b, \eval\,(\Loop\,a\,(\Arr\,{\mathit{swap}})))
    \end{array}
\]
It is worth noting that, in this example, the signal function produces a new signal function that is identical to the original,
except for an updated internal state. In fact, this property holds for all \Yampacore{} terms, where
multiple updates may occur. This characteristic significantly simplifies bisimulation proofs.
Finally, a term is lifted into a stream function by
applying the corecursive function \(\run\) to its stepwise semantics.
\begin{equation*}
    \begin{array}{lcll}
        \run &:& \sfType\,A\,B\rightarrow \StreamType\, A \rightarrow \StreamType\, B\\
        \run\, \sfval\, (a:s) &=& b:s'\\
        && {\text{where }} (b,\, \sfval') =
        \sfval\, a \land s' = \run\, \sfval'\, s
    \end{array}
    \label{eq:yampa-run}
\end{equation*}
The functions \(\arr\), \(\comp\), \(\first\) and \(\sloop\) define an arrow 
with loops on the
category \(\chost\). 
Furthermore, these functions are compatible with bisimilarity.
To establish these results, we first refine the definition of bisimulation specifically for signal functions.
A relation \(R\) is a bisimulation on \(\sfType\,A\,B\) if for all \(\sfvaln{1}\) and \(\sfvaln{2}\) such that 
\(R\,\sfvaln{1}\,\sfvaln{2}\):
\begin{center}
if \(\sfvaln{1}\,a = (b, \sfvaln{1}')\) then there exists \(\sfvaln{2}'\) such that 
\(\sfvaln{2}\,a = (b, \sfvaln{2}')\) and \(R\,\sfvaln{1}'\,\sfvaln{2}'\)
\end{center}
As usual, we define bisimilarity, denoted \(\cdot \sim \cdot \), as the largest bisimulation.
This relation is an equivalence relation, as stated in \cref{sec:preliminary}.
We now proceed to prove the two results mentioned above.
\begin{lemma}
    Let \(\sfval\), \(\sfvaln{1}\), \(\sfvaln{1}'\), \(\sfvaln{2}\) and \(\sfvaln{2}'\) be signal functions and let $v$ be a value. 
    The following properties hold:
    \begin{itemize}
        \item if \(\sfvaln{1} \sim \sfvaln{1}'\) and 
        \(\sfvaln{2} \sim \sfvaln{2}'\) then \(\comp\, \sfvaln{1}\, \sfvaln{2} \sim \comp\, \sfvaln{1}'\, \sfvaln{2}'\)
        \item if \(\sfval \sim \sfval'\) then \(\first\,\sfval \sim \first\,\sfval'\)
        \item if \(\sfval \sim \sfval'\) then \(\sloop\,v\,\sfval \sim \sloop\,v\,\sfval'\)
    \end{itemize}
    \label{lemma:bisim}
\end{lemma}
\begin{proof}
    The proof proceeds by bisimulation. For the first case, we define the relation
    \(R = \{ (\comp\, \sfvaln{1}\, \sfvaln{2},\,\comp\, \sgvaln{1}\, \sgvaln{2})\mid 
    \sfvaln{1} \sim \sgvaln{1} \wedge \sfvaln{2} \sim \sgvaln{2}\}\).
    To prove the result it is sufficient to prove that R is a bisimulation.
    Suppose that \(R\, \sfval\, \sgval\) and \(\sfval\,a = (b, \sfval')\).
    We need to prove that there exists \(\sgval'\) such that
    \(\sgval\, a = (b, \sgval')\) and \(R\,\sfval' \sgval'\) holds.
    By definition of \(R\), there exists \(\sfvaln{1}\), \(\sfvaln{2}\), 
    \(\sgvaln{1}\) and \(\sgvaln{2}\)
    such that \(\sfval = \comp\,\sfvaln{1}\,\sfvaln{2}\), \(\sgval = \comp\,\sgvaln{1}\,\sgvaln{2}\), 
        \(\sfvaln{1} \sim \sgvaln{1}\) and \(\sfvaln{2} \sim \sgvaln{2}\).
    By definition of \(\comp\), there exists \(c\), \(\sfvaln{1}'\) and 
    \(\sfvaln{2}'\) 
    such that
    \(\sfvaln{1}\, a = (c, \sfvaln{1}')\), \(\sfvaln{2}\, c = (b, \sfvaln{2}')\)
    and \(\sfval' = \comp\, \sfvaln{1}'\, \sfvaln{2}'\).    
    By 
    bisimilarity there exists \(\sgvaln{1}'\) and \(\sgvaln{2}'\) such that
    (1)~\(\sgvaln{1}\, a = (c, \sgvaln{1}')\) and 
    \(\sfvaln{1}' \sim \sgvaln{1}'\) and
    (2)~\(\sgvaln{2}\, c = (b, \sgvaln{2}')\) and \(\sfvaln{2}' \sim \sgvaln{2}'\).
    Thus taking \(\sgval' = \comp\, \sgvaln{1}'\, \sgvaln{2}'\), we have that
    \(\sgval\,a = (b, \sgval')\) and \(R\,\sfval'\,\sgval'\) holds.
    The other cases are similar, proving that both
    \(\{\first\,\sfval, \first\,\sgval \mid \sfval \sim \sgval\}\)
    and \(\{\sloop\,v\,\sfval, \sloop\,v\,\sgval \mid \sfval \sim \sgval\}\)
    are bisimulations.
\end{proof}
\begin{theorem}
    The functions \(\arr\), \({\mathit{comp}}\), \(\first\), and \({\mathit{loop}}\) 
    form an arrow with loops. 
    \label{prop:arrow}
\end{theorem}
\begin{proof}
    For example, to prove that \cref{eq:arrowlaw7} holds, we have to prove that
    \[\comp\, (\first\,(\first\,\sfval))\,(\arr\,{\mathit{assoc}}) \sim 
        comp\, (\arr\,{\mathit{assoc}})\,(\first\,\sfval)\]
    First, define the relation \[R_{ABC} = 
    \{(\comp\, (\first\,(\first\,\sfval))\,(\arr\,{\mathit{assoc}}),
    \comp\, (\arr\,{\mathit{assoc}})\,(\first\,\sfval))\}_{\sfval:\sfType\,A\,B\,C}\].    
    Next, suppose that \((\comp\, (\first\,(\first\,\sfval))\,(\arr\,{\mathit{assoc}}))\,((a_1,a_2), a_3) = 
    ((b,(a_2', a_3')),\sgval)\) for some \(b\), \(a_2'\), \(a_3'\) and \(\sgval\).
    By definition of the functions \(\comp\), \(\first\) and \(arr\) we have 
    \(a_2' = a_2\), \(a_3' = a_3\) and 
    \(\sfval\,a = (b,\sfval')\) and \(\sgval = \comp\, (\first\,(\first\,\sfval'))\,(\arr\,{\mathit{assoc}})\)
    for some \(\sfval'\).
    We also have \[(comp\, (\arr\,{\mathit{assoc}})\,(\first\,\sfval))\,((a_1,a_2), a_3) =
    ((b,(a_2, a_3)),\sgval')\] where 
    \(\sgval' = comp\, (\arr\,{\mathit{assoc}})\,(\first\,\sfval)\).
    By definition of \(R_{ABC}\), we have \(R_{ABC}\,\sgval\,\sgval'\).
    Other cases are similar. Take the 
    relation that connects the left and right sides of the equation,
    generalized over all variables. 
\end{proof}
As stated in \cref{sec:preliminary}, this arrow gives rise to
a category \(\csf\), where \(\arr\) is an identity-on-objects functor between \(\chost\) and \(\csf\). 
Together, these results enable us to establish the correctness of multiple transformations
in the following section.
\subsection{Normal form}
We demonstrate  that any \Yampacore{} program can be transformed into a normal form
\(\Loop\,v\,(\Arr\,f)\) for some value \(v\) and function \(f\) of the host language.
Building on bisimulation, we define the relation \(\equiv\) over \Yampacore{} terms as
\( t_1 \equiv t_2  {\text{ iff }}  \eval\,t_1 \sim \eval\,t_2 \). By \cref{lemma:bisim},
we can derive that \(\equiv\) is a congruence on the term algebras \(\SfType\,A\,B\) .
\begin{lemma} 
    Let \(t\), \(t_1\), \(t_1'\), \(t_2\) and \(t_2'\) be terms. The following properties hold:
    \begin{itemize}
        \item if \(t_1 \equiv t_1'\) and \(t_2 \equiv t_2'\) then \(\normalfont \Comp\,t_1\,t_2 \equiv \Comp\,t_1'\,t_2'\)
        \item if \(t \equiv t'\) then \(\normalfont \First\,t \equiv \First\,t'\)
        \item if \(t \equiv t'\) then \(\normalfont \Loop\,v\,t \equiv \Loop\,v\,t'\)
    \end{itemize}
    \label{lemma:congruence}
\end{lemma}
\begin{proof}
    Immediate by definition of $\equiv$ and \cref{lemma:bisim}.
\end{proof}
We now have all the necessary elements to prove that any \Yampacore{} term can be 
transformed into a normal form.
\begin{theorem}
    For every term \(t\), there exists a value \(v\) and a function \(f\) in
    the host language such that
    \(\normalfont t \equiv \Loop\,v\,(\Arr\,f)\)
    \label{prop:normalform}
\end{theorem}
\begin{proof}
    The proof is by structural induction on the term $t$ and by
    definition of \(\equiv\).
    \begin{itemize}
        \item \(\Arr\,f \equiv \Loop\,\void\,(\Arr\,(\lambda (x,y).(f\,x,\void)))\)
        \item if \(t \equiv \Loop\,v\,(\Arr\,f)\) then
        \(\First\,t \equiv \Loop\,v\,(\Arr\,g)\) where
        \( g = \lambda ((a,d),c).\letin{(b,c')}{f\,(a,c)}{((b,d),c')}\) 
        \item if \(t_1 \equiv \Loop\,d\,(\Arr\,f_1)\) and
        \(t_2 \equiv \Loop\,e\,(\Arr\,f_2)\) then
        \(\Comp\,t_1\,t_2 \equiv \Loop\,(d,e)\,(\Arr\,g)\)
        where 
        \(g = \lambda (a, (c_1,c_2)).
            \letin{(b,c_1')}{f_1\,(a,c_1)}{\letin{(c,c_2')}{(b,c_2)}{(c,(c_1',c_2'))}}\).
        \item if \(t \equiv \Loop\,d\,(\Arr\,f)\) then
        \(\Loop\,c\,t \equiv \Loop\,(c,d)\,(\Arr\,g)\) where
        \[g = \lambda (a,(c,d)).\letin{((b,c'), d')}{f\,((a,c),d)}{(b,(c',d'))}\]
    \end{itemize}
    In each case, 
    bisimilarity is established at the semantics level by considering the relation that connects the left and right sides
    of the equation, generalized over all variables, and ensuring that the 
    sub-expressions are bisimilar.
    For example, in the case of \(first\), take the relation
    \(R = \{ (\first\,\sfval, \sloop\,d\,(\arr\,g) ) 
            \mid \sfval \sim \sloop\,d\,(\arr\,f)\}\), where \(g\) is defined as above, 
    and prove that it is a bisimulation.
\end{proof}

%% file: pages/frp.tex
This section introduces \ourlanguage, a simple reactive functional language that extends \Wormholes.
Like \Wormholes, this language replaces the loop combinators with signal functions to read and write
resources. Unlike \Wormholes we do not introduce a binder for local resources, instead we rely on the status of resources
which are classified as inputs, outputs or internal
resources.
Input and output resources are single-use, meaning they must be accessed exactly once.
Internal resources, on the other hand, can be read from and written to at any time.
The language is equipped with a type system, presented in \cref{sec:static-semantics}, that ensures that input and output resources are used correctly.
\subsection{Syntax}
\label{sec:mh_syntax}
The syntax of \ourlanguage{} is presented in \cref{fig:whle-syntax}.
The type \(\RefType\, A\) represents resources that carry a value of type \(A\) and are uniquely
identified by a natural number. 
A value of type \(\RefType\, A\) is expressed as 
\(\Refcons\, A\, n\), where \(n\) is the resource identifier.
Terms constructed with \(\Arr\), \(\First\) and \(\Comp\) are similar to their counterparts in \Yampa.
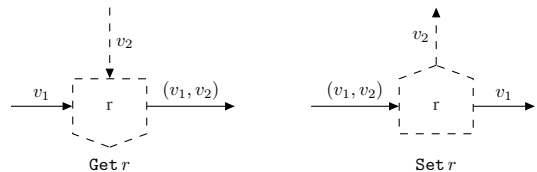
\begin{figure}[b]
    \centering
        
    \begin{tikzpicture}[scale=0.7]
  
      \node (getIS) at (-4,0) {};
      \node[get] (get) at (-2,0) {r};
      \node (getOS) at (0.5,0) {};
      \node (getE) at (-2,2) {};
      \node (getLegend) at (-2,-1.1) {$\Get\,r$};
  
      \draw[dfarrow] (getIS) to[edge label={$v_1$}, pos=0.5] (get);
      \draw[dfarrow, dashed] (getE) to[edge label={$v_2$}, pos=0.5] (get);
      \draw[dfarrow] (get)   to[edge label={$(v_1,v_2)$}, pos=0.5] (getOS);
  
      \node (setIS) at (1.7,0) {};
      \node[set] (set) at (4.2,0) {r};
      \node (setOS) at (6.2,0) {};
      \node (setE) at (4.2,2) {};
      \node (setLegend) at (4.2,-1.1) {$\Set\,r$};
  
      \draw[dfarrow] (setIS) to[edge label={$(v_1,v_2)$}, pos=0.5] (set);
      \draw[dfarrow, dashed] (set) to[edge label={$v_2$}, pos=0.5] (setE);
      \draw[dfarrow] (set)   to[edge label={$v_1$}, pos=0.5] (setOS);
    
    \end{tikzpicture}
    
    \caption{Graphical representation of resource accesses}
    \label{figure:resource:overview}
\end{figure}
The \(\Get\) and \(\Set\) constructs are used to read from and write to resources, respectively. A Graphical representation is provided in \cref{figure:resource:overview}.
\begin{itemize}
    \item \(\Get\, r\) represents a signal function that takes an input \(x\) and returns  a pair \((x,y)\), 
    where \(y\) is the value of the resource \(r\).
    \item \(\Set\, r\) represents a signal function that takes an input \((x,y)\) and returns \(x\), 
    after updating resource \(r\) with the value \(y\)
\end{itemize}

\begin{figure}[ht]
    \begin{equation*}
        \begin{array}{lcl}
            \RefType&=& \Refcons : \forall A.\, A \rightarrow \NatType \rightarrow \RefType\, A\\\\
            \RSfType &=& \mid \Arr : 
            \forall A\, B.\, 
            (A \rightarrow B) \rightarrow \RSfType\, A\, B\\
            && 
            \mid \First : \forall A\, B\, C.\,
             \RSfType\, A\, B \rightarrow \RSfType\, (A \times C)\, (B \times C)\\
            &&\mid  
            \Comp : \forall A\, B\, C.\,
            \RSfType \, A\, B \rightarrow \RSfType\, B\, C \rightarrow \RSfType\, A\, C\\
            &&\mid \Get : \forall A\, B.\, \RefType\, B \rightarrow \RSfType \, A\, (A \times B)\\ 
            &&\mid \Set : \forall A\, B.\, \RefType\, B \rightarrow \RSfType\, (A \times B)\, A
        \end{array}
    \end{equation*}
    \caption{Syntax of \ourlanguage{}}
    \label{fig:whle-syntax}
    \end{figure}
We adopt a programming model in which all interactions with the environment are mediated through resources. Consequently, both input and output types of entire programs are restricted to the singleton type \(\VoidType\),
whose only value is denoted as \({\tt{tt}}\).
A program is defined by a term of type \(\RSfType\,\VoidType\,\VoidType\) along with mappings that associate its 
input and output resources with types and its internal resources with initial typed values.
\[
\begin{array}{lcll}
    \ProgType &=& \{ & 
    {\tt{inputs}} : \ListType\,\TypeType;\\
    &&&{\tt{outputs}} : \ListType\,\TypeType;\\
    &&&{\tt{internals}} : \ListType\,Val_\TypeType;\\
    &&&{\tt{program}} : \RSfType\, \VoidType\, \VoidType \, \}
\end{array}
\]
where \(\TypeType\) represents the types of the host language and \(Val_\TypeType\) represents typed values
in the host language.
\begin{example}
    The delay function presented in \cref{sec:preliminary} can be expressed in \ourlanguage{} as follows:
    \[
    (\Get\, r) >>> \Arr\, (\lambda (x,y).(y,x)) >>> \Set\, r : \RSfType\, A\, A
    \]
    where for \(r : \RefType\, A\), the first value of the output stream is the initial value 
    of \(r\).
    A complete example is the program below, which generates the sequence of natural numbers:
    \[
        \begin{array}{ll}
    \{&
        {\tt{inputs}}=[]; \\
          &  {\tt{outputs}}=[{\mathit{Nat}}];\\
           & {\tt{internals}}=[0:{\mathit{Nat}}];\\
           & {\tt{program}} =     
                \Get\, r_0 >>> \Arr\, (\lambda ((),n) . (((),n),n+1)) >>> \Set\, r_0 >>> \Set\,r_1
    \}
        \end{array}
    \]
    where $r$ is the unique input resource and $r_0$ is the unique internal resource.
\end{example}


\subsection{Dynamic semantics}
\label{sec:mhsemantics}
The semantic domain of stepwise computations in \ourlanguage{} is the Kleisli category associated with 
a state monad that manages memory. This memory accounts for different types of resources and their access rights,
making the semantics  a partial function. 
However, the type system of \cref{sec:static-semantics} ensures 
that this function is total for well-typed programs.
The memory domain and the monad built on top of it are introduced in \cref{sec:mhmemory} and \cref{sec:mhstate}, respectively.
The stepwise semantics domain is formalized in \cref{sec:mhsemanticdomain} as the Kleisli category associated with the state monad.
Similarly to what we have done for \Yampacore{}, we demonstrate that this semantic domain forms an arrow.
We also introduce a set of equations satisfied by the new constructors, \(\Get\) and \(\Set\).
These equations serve as the counterpart to the laws governing loops in \Yampacore{}.
Finally, the stream-based semantics of programs is formalized in \cref{sec:mhstream}.
\subsubsection{Memory}
\label{sec:mhmemory}
    Memory is represented as elements of type \(\MemoryType\), defined as partial mappings (denoted by \(\rightharpoonup\)) from resource identifiers to cells (see \cref{fig:memory}). 
    A cell, of type \(\CellType\), is a pair consisting of a tag of type \(\TagType\) and an optional value of the type specified by the tag. Given a type \(A\) we denote by \(A^?\) the type \(A\) extended with an additional element \({\tt{undef}}\).
    The tag itself is a pair containing a status, which indicates the accessibility of the resource, and a 
    type, which specifies the type of values carried by the resource.
    The status of a resource can take one of the following forms : 
    \begin{itemize}
        \item \(\Internal\): Represents an internal resource, which can be read from and written to at any time.
        \item \(\Input\, b\): Represents an input resource, which can only be read from if the boolean \(b\) is true.
        \item \(\Output\, b\): Represents an output resource, which can only be written if the boolean \(b\) is true.
    \end{itemize}
    \begin{figure}
    \begin{equation*}
        \begin{array}{l}
        \begin{array}{lcl}
            \StatusType &=&
            \mid \Internal : \StatusType \\
            &&\mid \Input : \BoolType \rightarrow \StatusType\\
            &&\mid \Output : \BoolType \rightarrow \StatusType
            \\
            \TagType &=& \StatusType \times \TypeType
            \\
            \CellType &=&
            \Cell : \forall t :\TagType. 
            ({\mathit{snd}}\,t)^? \rightarrow \CellType\\
            \MemoryType &=& \NatType \rightharpoonup \CellType
        \end{array}
        \\\\
        \begin{array}{lcll}
            \mread &:& \RefType\, A \rightarrow \MemoryType \rightharpoonup  A \times \MemoryType\\
            \mread\, r\, \sigma &=& 
            {\tt{case}}\, \sigma\, (\id \, r)\, {\tt{of}}\\
            && \mid \Cell\, (\Internal, B)\, x\,  \Rightarrow (x,\sigma) & {\text{if }} A = B\\
            &&\mid \Cell\, (\Input\, \true, B)\, x\, \\
                && \qquad \Rightarrow (x,\sigma[(\id \, r) \mapsto \Cell\, (\Input\, \false, B)\, {\tt{undef}}])
                & {\text{if }} A = B\\
            \\\\
            \mwrite &:& \RefType\, A \rightarrow \MemoryType \rightarrow  A \rightharpoonup \MemoryType\\
            \mwrite\, r\, \sigma\, v &=&
            {\tt{case}}\, \sigma\,(\id\, r) \, {\tt{of}}\\
            &&\mid \Cell\, (\Internal, B)\, x\,  \\
            && \qquad \Rightarrow \sigma[(\id \, r) \mapsto \Cell\, (\Internal, B)\, v] & {\text{if }} A = B\\
            &&\mid \Cell\, (\Output\, \true, B)\, {\tt{undef}}\, \\
            && \qquad \Rightarrow \sigma[(\id \, r) \mapsto \Cell\, (\Output\, \false, B)\, v]& {\text{if }} A = B
        \end{array}\\
        {\text{where }} \sigma[n \mapsto c] = \lambda n'.\, 
        {\tt{if }}\, n = n'\, {\tt{ then }}\,\,c\,\,{\tt{ else }}\,\, \sigma\, n'
    \end{array}
    \end{equation*}
    \caption{Memory domain}
    \label{fig:memory}
\end{figure}
The predicates \(\readable\) and \(\writable\) determine
whether a resource is readable or writable within a given memory. 
The functions \(\mread\, r\) and \(\mwrite\, r\, v\) are total over memories satisfying 
\(\readable\, r\) and \(\writable\, r\), respectively.
\begin{equation*}
    \begin{array}{lcl}
        \readable\, (\Refcons{}\, A\, n)\, \sigma &=& 
        (\exists v:A.\, \sigma\, n = \Cell\, (\Internal, A)\, v) \lor\\
        && \qquad (\exists v:A.\, \sigma\, n = \Cell\, (\Input\, \true, A)\, v)
        \\
        \writable\, (\Refcons{}\, A\, n)\, \sigma &=& 
        (\exists v:A.\, \sigma\, n = \Cell\, (\Internal, A)\, v) \lor\\
        && \qquad (\sigma\, n = \Cell\, (\Output\, \true, A)\, {\tt{undef}})
    \end{array}
    \label{eq:readable-writable}
\end{equation*}
Note that in the definitions of \(\mread\) and \(\mwrite\) we require that
the type of the value stored in the cell matches the type of the resource.
This may not always be the case, as the memory is only aware of the identifier of the
resource and not its type. For well-typed programs, the two types are always identical.
\subsubsection{State Monad}
\label{sec:mhstate}
    Building on the type \(\MemoryType\), we define a state monad based on the functor
    \[
        \begin{array}{lcl}
            \State\, A &=& \MemoryType \rightharpoonup A \times \MemoryType\\
            \State\,f &=& \lambda h\,\sigma.(f\,a,\sigma') {\text{ where }} (a,\sigma') = h\, \sigma
        \end{array}
    \]
    This monad is equipped with the following operations:
    \begin{equation*}
        \begin{array}{lclclcl}
            \Stateret_A &::& A \rightarrow \State\, A &&
            \Statebind_{AB} &::& \State\, A \rightarrow (A \rightarrow \State\, B) \rightarrow \State\, B\\
            \Stateret\, a &=& \lambda \sigma. (a,\sigma)&&
            \Statebind\, m\, f &=& \lambda \sigma. f\, x\, \sigma' \text{ where } (x,\sigma') = m\, \sigma
            \\\\
            \Stateget_A\, r &::& \RefType\,A \rightarrow \State\, A&&
            \Stateset_A\, r\, v &::& \RefType\, A \rightarrow A \rightarrow \State\, \VoidType\\
            \Stateget\, r &=& \mread\,r&&
            \Stateset\, r\,v &=& \lambda \sigma. (\void, \mwrite\,r\,\sigma\,v)
        \end{array} 
    \end{equation*}
    The functions \(\Stateget\, r\) and \(\Stateset\,r\,v\) are total over memories satisfying the predicates \(\readable\, r\) and \(\writable\, r\), respectively.
    These operations satisfy the monad laws presented in \cref{eq:monadlaw1,eq:monadlaw2,eq:monadlaw3}.
    We extend this equational theory with laws that establish relationships between the \(\Stateget\) and \(\Stateset\) operations.
    The validity of these laws for the state monad follows directly from the definition of the memory domain.
    Only internal resources are concerned by 
    \cref{eq:statemonadlaw1,eq:statemonadlaw2,eq:statemonadlaw3,eq:statemonadlaw4,eq:statemonadlaw5}.
    These rules allow the removal of certain resources accesses, which is not permitted for input and output resources, as
    their accesses update their status (readable or writable). 
    Consequently, equations \cref{eq:statemonadlaw1,eq:statemonadlaw2,eq:statemonadlaw3,eq:statemonadlaw4,eq:statemonadlaw5} should be interpreted as follows: \(f = g\) means \(f\,\sigma = g\,\sigma\) for all \(\sigma\) where
    the resource \(r\) is an internal resource.
    Similarly, other rules apply to all types of resources, provided that for read (resp. write) access, the resource is readable
    (resp. writable).
    \begin{subequations}\label{eq:statemonadlawsgs}
        \begin{align}
            \Statebind\, (\Stateget\, r)\, (\lambda x.\Stateget\, r) &= \Stateget\, r\label{eq:statemonadlaw1}\\
            \Statebind\, (\Stateget\, r)\, (\Stateset\, r) &= \Stateret\, {\tt{tt}}\label{eq:statemonadlaw2}\\
            \Statebind\, (\Stateset\, r\, x)\, (\lambda \void.\, \Stateget\, r) 
            &= \Statebind\, (\Stateset\, r\, x)\, (\lambda \void.\, \Stateret\, x)\label{eq:statemonadlaw3}\\
            \Statebind\, (\Stateset\, r\, x)\, (\lambda {\tt{tt}}.\Stateset\,r\,y) &= 
            \Stateset\, r\, y\label{eq:statemonadlaw4}\\
            \Statebind\, (\Stateget\,r)\, (\lambda y. \Stateret\,x) &= \Stateret\,x
            \label{eq:statemonadlaw5}\\
            \Statebind\, (\Stateget\, r)\, (\lambda x.\Stateget\, r') &=
            \Statebind\, (\Stateget\, r')\, 
                (\lambda x. \Statebind\, (\Stateget\,r)\, (\lambda y. \Stateret\,x))
                \label{eq:statemonadlaw6}
            \\
            \Statebind\, (\Stateget\, r)\, (\lambda x.\Stateset\, r'\,y) &=
            \Statebind\, (\Stateset\, r'\,y)\, 
            (\lambda \void.\Statebind\,(\Stateget\, r)\,(\lambda x.\Stateret\,{\tt{tt}}))
            \label{eq:statemonadlaw7}
            \\
            \Statebind\, (\Stateset\, r\,x)\, (\lambda {\tt{tt}}.\Stateget\, r') &=
            \Statebind\, (\Stateget\, r')\, (\lambda y. \Statebind\,(\Stateset\,r\, x)\, (\lambda \void.\Stateret\, y))
            \label{eq:statemonadlaw8}
            \\
            \Statebind\, (\Stateset\, r\,x)\, (\lambda {\tt{tt}}.\Stateset\, r'\,y) &=
            \Statebind\, (\Stateset\, r'\,y)\, (\lambda {\tt{tt}}.\Stateset\, r\,x)  
            \label{eq:statemonadlaw9}
        \end{align}
    \end{subequations}
\begin{lemma}
    The functor \(\State\), along with the operations defined above, satisfies the monad laws
    presented in {\normalfont \cref{eq:monadlaw1,eq:monadlaw2,eq:monadlaw3}} and the additional laws given in
    {\normalfont \cref{eq:statemonadlaw1,eq:statemonadlaw2,eq:statemonadlaw3,eq:statemonadlaw4,eq:statemonadlaw6,eq:statemonadlaw7,eq:statemonadlaw8,eq:statemonadlaw9}}.
\end{lemma}
\begin{proof}
    The proof follows directly from the definitions of the operators and the memory domain.
    In each case, expanding the definition reduces the proof to a property of the memory domain.
    For instance, for \cref{eq:statemonadlaw2}, assuming the resource is internal
    we must show that if \(\mread\,r\,\sigma = (a, \sigma')\) then \(\sigma' = \sigma\)
    and \(\mwrite\,r\,\sigma\,a = \sigma\). This holds because accesses to an internal resource do
    not affect access rights.
    Other cases follow similarly.
\end{proof}
\begin{example}
By combining these equations, several equivalences can be derived.
For example, applying \cref{eq:monadlaw2}, \cref{eq:statemonadlaw5} and \cref{eq:statemonadlaw6} to an internal resource \(r\) allows the removal of an unused read access.
$$
\begin{array}{lcl}
    \Statebind\, (\Stateget\, r)\, (\lambda x.\Stateget\, r') &=& 
            \Statebind\, (\Stateget\, r')\, 
                (\lambda x. \Statebind\, (\Stateget\,r)\, (\lambda y. \Stateret\,x))\\
    &=& \Statebind\, (\Stateget\, r')\, \Stateret\\
    &=& \Stateget\, r'
\end{array}
$$
\end{example}
\subsection{Semantic domain}
\label{sec:mhsemanticdomain}
    Now that we have all the necessary elements, we can define the semantic domain of \ourlanguage{} and
    prove that it satisfies the arrow laws, as well as additional laws similar to those governing loops
    in \Yampacore{}. Together, these laws enable us to prove the correctness of the transformations
    presented in the next section.
    The semantic domain of combinators is defined by the type 
    \[\rsfType\, A\, B = A \rightarrow \State\, B\]
    Notably, unlike the definition of the semantic domains \(\sfType\,A\,B\), the definition of
    \(\rsfType\, A\, B\) is not corecursive.
    As a result, program transformations within \ourlanguage{} are greatly simplified.
    Only the final transformation to \Yampa{} will require the use of bisimulation.
    To achieve this we lift this semantics domain to a coalgebra of the functors \(F_{A,B}\).
    
    Building on this definition, we define the category \(\crsf\) as the Kleisli category associated
    with the monad \(\State\). This category has the types of the host language as objects and morphisms 
    of type \(\rsfType\, A\, B\). As stated in \cref{sec:preliminary}, identity morphisms and compositions are given by 
    \(\Stateret\) and \(\Statebind\), respectively. Their definitions are restated below.
    \[
    \begin{array}{lcl}
        id &=& \lambda x.\, \Stateret\, x\\
        (g \circ f) &=& \lambda x.\, \Statebind\, (f\, x)\, g
    \end{array}
    \]
    Next, we define the following functions, where the \(arr\) function defines 
    identity-on-objects functors between \(\chost\) and \(\crsf\).
    \begin{equation*}
        \begin{array}{lcl}
        \arr\, f &=& \lambda x.\, \Stateret\, (f\, x)\\
        \comp\,\rsfvaln{1}\,\rsfvaln{2}&=& \rsfvaln{2} \circ \rsfvaln{1}\\
        \first\, \rsfval &=& \lambda (x,c).\, \Statebind\, (\rsfval\, x)\, 
        (\lambda y.\, \Stateret\, (y,c))\\
        \get\,r &=& \lambda x.\, \Statebind\, (\Stateget\, r)\, 
        (\lambda y.\, \Stateret\, (x,y))\\
        \set\,r &=& \lambda (x,y).\, \Statebind\, (\Stateset\, r\, y)\, 
        (\lambda \void.\, \Stateret\, x)
        \end{array}
    \end{equation*}
\begin{theorem}
    The functions $\arr$, $\first$, and $\comp$ form an arrow.
\end{theorem}
\begin{proof}
    The proof follows by simple equational reasoning. We prove that \cref{eq:arrowlaw4} holds.
    By definition of \(arr\) and \(comp\), we have \(arr\,(g \circ f) = \lambda x.\Stateret\,(g\,(f\,x))\) and
    \(\arr\,f \circ \arr\,g = \lambda x.\Statebind\,(\Stateret\,(f\,x))\,(\arr\,g)\).
    Applying \cref{eq:monadlaw1} in the right-hand side, we obtain the left-hand side.
    Other cases are similar, using the monad laws for \(\State\).
 \end{proof}
In addition to arrow laws, the \(\get\) and \(set\) function bring new equations defined in \cref{eq:rewriteafc:get,eq:rewriteafc:set,eq:rewriteafc:getset}.
Those equations play a role similar to the laws governing loops in \Yampacore{} but are separated
into three groups. 
The first group governs the reordering of \(get\) operations, allowing them to be moved to the left or eliminated. 
\begin{subequations}\label{eq:rewriteafc:get}
    \begin{align}
        \arr\,f \compop \get\,r &= \get\,r \compop \first\,\arr\,f\label{eq:rewriteafc1} \\
        \first\,(\get\,r \compop \rsfval) &= \get\,r \compop \arr\,\perm \compop \first\,\rsfval\label{eq:rewriteafc3} \\
        \first\,(\rsfval) \compop \get\,r &= \first\,(\rsfval \compop \get\,r) \compop \arr\,\perm\label{eq:rewriteafc5} \\
        \get\,r \compop \get\,r' &= \get\,r' \compop \get\,r \compop \arr\,\perm\label{eq:rewriteafc8}\\
        \get\,r \compop \get\,r &= \get\,r \compop \arr\,\sdup\label{eq:rewriteafc7}
    \end{align}
\end{subequations}
Similarly, the second groups governs the reordering of \(set\) operations, allowing them to be moved to the right or eliminated.
\begin{subequations}\label{eq:rewriteafc:set}
    \begin{align}
        \set\,r \compop \arr\,f &= \first\,\arr\,f \compop \set\,r\label{eq:rewriteafc2} \\
        \first\,(\rsfval \compop \set\,r) &= \first\,\rsfval \compop \arr\,\perm \compop \set\,r\label{eq:rewriteafc4} \\
        \set\,r \compop \first\,(\rsfval) &= \arr\,\perm \compop \first\,(\set\,r \compop \rsfval)\label{eq:rewriteafc6} \\
        \set\,r \compop \set\,r' &= \arr\,\perm \compop \set\,r' \compop \set\,r \label{eq:rewriteafc10}\\
        \set\,r \compop \set\,r &= \arr\,\fst \compop \set\,r\label{eq:rewriteafc9}
    \end{align}
\end{subequations}
Finally, the third group governs the reordering of consecutive \(set\) and \(get\) operations. 
\cref{eq:rewriteafc11,eq:rewriteafc12} apply only when the resource is internal, 
while \cref{eq:rewriteafc13} holds for any resource.
\begin{subequations}\label{eq:rewriteafc:getset}
    \begin{align}
        \set\,r \compop \get\,r' &= \get\,r' \compop \arr\,\perm \compop \set\,r \label{eq:rewriteafc13}\\
        \set\,r \compop \get\,r &= \arr\,\sdup \compop \set\,r \label{eq:rewriteafc11}\\
        \get\,r \compop \set\,r &= \arr\,id \label{eq:rewriteafc12}
    \end{align}
\end{subequations}
Those laws are satisfied, thus forming an arrow with resources that mimic arrows with loops.
\begin{lemma}{}
{\normalfont \cref{eq:rewriteafc1,eq:rewriteafc3,eq:rewriteafc5,eq:rewriteafc8}},
{\normalfont \cref{eq:rewriteafc2,eq:rewriteafc4,eq:rewriteafc6,eq:rewriteafc10}}
and {\normalfont \cref{eq:rewriteafc13}} hold for any resources.
{\normalfont \cref{eq:rewriteafc7,eq:rewriteafc9,eq:rewriteafc11,eq:rewriteafc12}} hold for internal resources.
\end{lemma}
\begin{proof}
    The proof follows by simple equational reasoning, using the definitions, the monad laws 
    \cref{eq:monadlaw1,eq:monadlaw2,eq:monadlaw3}, 
    and the state monad laws 
    \cref{eq:statemonadlaw1,eq:statemonadlaw2,eq:statemonadlaw3,eq:statemonadlaw4,eq:statemonadlaw5,eq:statemonadlaw6,eq:statemonadlaw7,eq:statemonadlaw8,eq:statemonadlaw9}.
    We prove that \cref{eq:rewriteafc1} holds,
    other cases follows similarly.  
    \[
    \begin{array}{lll}
        & \get\,r \compop \first\,\arr\,f &\\
        =& \lambda x.\Statebind\,(\Statebind\,(\Stateget\,r)\,(\lambda y.\Stateret\,(x,y)))\,(\first\,\arr\,f) & ({\mathit{def}}.)\\
        =& \lambda x.\Statebind\,(\Stateget\,r)\,(\lambda y.\Statebind\,(\Stateret\,(x,y))\,(\first\,\arr\,f)) & (\cref{eq:monadlaw3})\\
        =& \lambda x.\Statebind\,(\Stateget\,r)\,(\lambda y.(\first\,\arr\,f)\,(x,y)) & (\cref{eq:monadlaw1})\\
        =& \lambda x.\Statebind\,(\Stateget\,r)\,(\lambda y.(\Statebind\,(\Stateret\,(f\,x))\,(\lambda z.\Stateret (z,y)))) & ({\mathit{def}}.)\\
        =& \lambda x.\Statebind\,(\Stateget\,r)\,(\lambda y.(\Stateret\,(f\,x,y))) & (\cref{eq:monadlaw1})\\
        =& \lambda x.\get\,r\,(f\,x)& ({\mathit{def}}.)\\
        =& \lambda x.\Statebind\,(\Stateret (f\,x))\,\get\,r& (\cref{eq:monadlaw1}) \\
        =& \arr\,f \compop \get\,r& ({\mathit{def}}.) 
    \end{array}
\]  
\end{proof}
\subsubsection{Stepwise evaluation}
\label{sec:mhstepwise}
The functions defined previously enable a simple recursive definition of the stepwise semantics
of \ourlanguage{} terms. This semantics maps a term of type \(\RSfType\, A\, B\) to a morphism in \(\crsf\).
\begin{equation*}
\begin{array}{lcl}
    \eval &::& \RSfType\, A\, B \rightarrow \rsfType\, A
    \, B\\
    \eval\, (\Arr\, f) &=& \arr\, f\\
    \eval\, (\Comp\, \rsfvaln{1}\, \rsfvaln{2}) &=& \comp\,(\eval\, \rsfvaln{1})\,(\eval\, \rsfvaln{2})\\
    \eval\, (\First\, \rsfval) &=& \first\, (\eval\, \rsfval)\\
    \eval\, (\Get\, r) &=& \get\,r\\
    \eval\, (\Set\, r) &=& \set\,r
\end{array}
\label{eq:eval}
\end{equation*}
The relation \(\equiv\) over \ourlanguage{} terms is defined as
\( t \equiv t' {\text{ iff }} \eval\, t = \eval\, t'\).
This relation is a congruence, as stated by the following straightforward lemma.
\begin{lemma}\label{lemma:equiv:Molholes:congruence}
Let \(t\), \(t_1\), \(t_1'\), \(t_2\), \(t_2'\) be terms of \ourlanguage{}.
The following properties hold:
\begin{itemize}
    \item if \(t_1 \equiv t_1'\) and \(t_2 \equiv t_2'\) then \(\normalfont \Comp\,t_1\, t_2 \equiv \Comp\,t_1'\,t_2'\)
    \item if \(t\equiv t'\) then \(\normalfont \First\,\,t \equiv \First\,\,t'\)
\end{itemize}
\end{lemma}
\begin{proof}
    This follows immediately from the definition of \(\equiv\).
\end{proof}
\subsubsection{Stream functions}
\label{sec:mhstream}
The semantics of a program $p$ has type 
\(\StreamType\,({\tt{inputs}}\,p) \rightarrow \StreamType\, ({\tt{outputs}}\,p)\).
It is given by 
${\tt{run}}\,p\,(\init\, p)$ where {\tt{run}} is defined as follows:
\begin{equation*}
    \begin{array}{lcl}
    \run &:& \ProgType \rightarrow \MemoryType \rightarrow \StreamType\,({\tt{inputs}}\,\,p) 
    \rightharpoonup \StreamType\,({\tt{outputs}}\,p)
    \\    \run\,p\,\sigma\,(i:s) &=& (\push\,p\,\sigma') : \run\,p\, \sigma'\, s\\
    && {\text{where }}
    ({\tt{tt}},\sigma') = 
    \eval\,({\tt{program\,p}})\, {\tt{tt}}\,(\pull\, p\,\sigma\,i)
    \end{array}
\end{equation*}
where, noting \(k\, p = |{\tt{internals}}\,p|\), 
\(k_{in}\,p = |{\tt{inputs}}\,p|\) and \(k_{out}\,p = |{\tt{outputs}}\,p|\),
the functions \(\init\), \(\pull\) and \(\push\) are defined as follows:
\[
\begin{array}{lcl}
    \init &:& \ProgType \rightarrow \MemoryType\\
    \init\, p\, &=& 
    \begin{cases} 
        \Cell\, (\Internal, \tau)\, v & {\text{if }}
        k_{in}\,p \leq n < k_{in}\,p+k\,p\\& {\text{and }} v:\tau = ({\tt{internals}}\, p)_n\\
        {\text{undefined}} & \text{otherwise}
    \end{cases}
    \\\\
    \pull &:& \ProgType \rightarrow \MemoryType \rightarrow \ListType\, \ValType \rightarrow \MemoryType\\
    \pull\, p\,\sigma\,i\,n &=&
    \begin{cases} 
        \Cell\, (\Input\,{\tt{true}}, \tau)\, v & {\text{if }}
        n < k_{in}\,p,\,
        \tau = ({\tt{inputs}}\, p)_n\\& {\text{and }} v:\tau = i_n\\
        \Cell\, (\Output\,{\tt{true}}, \tau)\, {\tt{undef}} & {\text{if }}
        k_{in}\,p + k_p \leq n < k_{in}\,p + k\,p + k_{out}\,p\\
        & {\text{and }} \tau = ({\tt{outputs}}\, p)_n\\
        \sigma\, n & \text{otherwise}
        \end{cases}
    \\\\
    \push &:& \ProgType \rightarrow \MemoryType \rightarrow \ListType\, \ValType\\
    \push\,p\,\sigma &=&
    \left [v:\tau 
        \middle | 
        \begin{array}{l}
            k_{in}\,p + k\,p \leq n < k_{in}\,p + k_p + k_{out}\,p\\
            \qquad {\text{and }} \sigma\,n= \Cell\, (\Output\,{\tt{false}}, \tau)\, v
        \end{array}
            \right ]
\end{array}
\]
The function \({\mathit{run}}\) is partial only because the function \(\eval\) is partial.
Therefore,
it is total for well-typed programs, as defined in the next section.
\subsection{Static semantics}
\label{sec:static-semantics}
\input{pages/frp_static.tex}

%% file: pages/frp_static.tex
We define a simple static semantics for \ourlanguage{}, ensuring the correct use of memory. Specifically, it guarantees that in well-typed programs, each input resource is read exactly once and each output resource is written exactly once.
\subsubsection{Abstract domain}
The abstract domain consists of an abstract memory, which maps resource identifiers to
abstract memory cells. These memory cells are simply tags, as defined in the previous section.
Models of an abstract memory \(\Sigma\) correspond to the set \(\gamma\,\Sigma\)
of concrete memories that share the same domain as \(\Sigma\) and where
the corresponding concrete and abstract cells are also related.
The abstraction relation over memory cells imposes that the tags is the same
and that the value held by the concrete memory has the correct type when the tag 
indicates that it must be defined.
\begin{equation*}
    \begin{array}{lcl}
    \AMemoryType = \NatType \rightharpoonup \TagType\\\\
    \gamma\,(st,\tau) = 
    \{ \Cell\,(st,\tau)\,v \mid 
        {\text{if }} st \in {\tt{defined}}
        {\text{ then }} v {\text{ is a value of type }} \tau
     \}\\
     {\text{where }} {\tt{defined}}= \{\Internal,\Input\,{\tt{true}},\Output\,{\tt{false}}\}\\\\
    \gamma\,\Sigma = \{ \sigma \mid \domain(\sigma) = \domain(\Sigma) \wedge 
        \forall n. n \in \domain(\sigma) \rightarrow \sigma\,n \in \gamma\,(\Sigma\,n)\}
    \end{array}
\end{equation*}
We define abstract \(\aread\) and \(\awrite\) operations, along with the predicates \(\areadable\) and \(\awritable\), which ensure that these operations are well-defined.
These operations update the status of memory cells, reflecting the fact that access has occurred.
\begin{equation*}
    \begin{array}{lcl}
        \aread &:& \RefType\, A \rightarrow \AMemoryType  \rightharpoonup \AMemoryType\\
        \aread\, r\,\Sigma &=& 
        {\tt{case}}\, \Sigma\, (\id \, r)\, {\tt{of}}\\
        && \mid (\Internal, B)  \Rightarrow \Sigma\\
        &&\mid (\Input\, \true, B) \Rightarrow \Sigma\,[\id \, r \mapsto (\Input\, \false, B)]
        \\\\
        \awrite &:& \RefType\, A \rightarrow \AMemoryType \rightharpoonup \AMemoryType\\
        \awrite\, r\,\Sigma &=&
        {\tt{case}}\, \Sigma\,(\id\, r) \, {\tt{of}}\\
        &&\mid (\Internal, B) \Rightarrow \Sigma\\
        &&\mid (\Output\, \true, B) \Rightarrow \Sigma[\id \, r \mapsto (\Output\, \false, B)]\\\\
        \areadable\, (\Refcons{}\, A\, n)\, \Sigma &=& \Sigma\, n = (\Internal, A) \lor \Sigma\, n = (\Input\, \true, A)\\
        \awritable\, (\Refcons{}\, A\, n)\, \Sigma &=& \Sigma\, n = (\Internal, A) \lor \Sigma\, n = (\Output\, \true, A)
    \end{array}
\label{eq:abstract-memory}
\end{equation*}
Note that, as with the definition of \(\mread\) and \(\mwrite\) in the previous section,
the type contained in a cell may differ from the type of the resource. 
Given a resource \(r:\RefType\,A\) and an abstract memory \(\Sigma\), we denote \(r \vdash \Sigma\) if the type of the tag associated 
with \(r\) in \(\Sigma\) is \(A\).
The abstraction relation defined by \(\gamma\) ensures that types remain consistent in 
a concrete memory, provided they are consistent in an abstraction of it.
The following lemma establishes that resource accesses preserve abstractions.
\begin{lemma}
    Let \(r\) be a resource of type \(\RefType\,A\), \(\sigma\) be a memory and
    let \(\Sigma\)
    be an abstract memory such that \(r \vdash \Sigma\): 
    \begin{itemize} 
    \item  if \(\sigma \in \gamma\,\Sigma\)
    and \(\normalfont \aread\,r\,\Sigma = \Sigma'\), then there exists a value \(a\) of type \(A\) and a memory \(\sigma'\) such that
    \(\normalfont \mread\,r\,\sigma = (a,\sigma')\), \(\sigma' \in \gamma\,\Sigma'\) and \(r \vdash \Sigma'\). Moreover, for all resource \(r' \not = r\), \(\Sigma \, r' = \Sigma \,r\).
    \item if \(\sigma \in \gamma\,\Sigma\)
    and \(\normalfont \awrite\,r\,\Sigma = \Sigma'\), then, for any value \(a\) of type \(A\), there  
    exists a memory \(\sigma'\) such that
    \(\normalfont \mwrite\,r\,\sigma\,a = \sigma'\), \(\sigma' \in \gamma\,\Sigma'\) and \(r \vdash \Sigma'\). Moreover, for all resource \(r' \not = r\), \(\Sigma \, r' = \Sigma \,r\).
    \end{itemize}
    \label{lemma:res_preservation}
\end{lemma}
\begin{proof}
Immediate by definition of \(\mread\), \(\mwrite\), \(\aread\) and \(\awrite\).
\end{proof}
\subsubsection{Abstract semantics}
The abstract semantics of a term is defined as a partial function over 
abstract memories.
For terms of the form \(\Arr\,f\), the abstract memory remains unchanged.
For terms of the form \(\Get\,r\) and \(\Set\,r\), the abstract semantics 
updates the state of memory cells using \(\aread\) and \(\awrite\) operations.
For terms of the form \(\First\) or \(\Comp\), changes in the abstract memory propagate.
\begin{equation*}
    \begin{array}{lcll}
        \aeval &::& &\RSfType \rightarrow \AMemoryType \rightharpoonup \AMemoryType\\
        \aeval\, (\Arr\, f)\, \Sigma &=& \Sigma\\
        \aeval\, (\First\, \rsfval)\, \Sigma &=& \Sigma' & \text{ if } \aeval\, \rsfval\, \Sigma =\Sigma'\\
        \aeval\, (\Comp\, \rsfvaln{1}\, \rsfvaln{2})\, \Sigma &=& \Sigma' & \text{ if } \aeval\, \rsfvaln{1}\, \Sigma = \Sigma'' \land \aeval\, \rsfvaln{2}\, \Sigma'' = \Sigma'\\
        \aeval\, (\Get\, r)\, \Sigma &=& \Sigma' & \text{ if } \areadable\,r\, \Sigma \land \aread\,r\, \Sigma = \Sigma'\\
        \aeval\, (\Set\, r)\, \Sigma &=& \Sigma' & \text{ if } \awritable\, r\,\Sigma \land \awrite\, r\,\Sigma = \Sigma'
\end{array}
\end{equation*}
A program \(p\) is well-typed if its abstract semantics is defined over an initial abstract memory given by \({\ainit}\,p\), where all inputs are readable (\(\Input\,{\tt{true}}\)) and all outputs are writable (\(\Output\,{\tt{true}}\)). 
Furthermore, we impose an additional condition that in the resulting abstract memory, 
all inputs must have the tag \(\Input\,{\tt{false}}\) and all outputs must have the tag
\(\Output\,{\tt{false}}\). 
The condition on output resources is mandatory to ensure that 
the program produces outputs at every step. The condition over input resources is optional, 
but useful to ensure that the program consumes all its inputs.
Below we define the \(\ainit\) functions and the well-typedness condition.
\begin{equation*}
    \begin{array}{lcl}
        \ainit &:& \ProgType \rightarrow \AMemoryType\\
        \ainit\, p\,n&=& 
        \begin{cases}
            \Cell\,(\Input\,{\tt{true}},\tau) & \text{if } n < k_{in}\, {~\text{and}~} 
            \tau = ({\tt{inputs\,p}})_n\\
            \Cell\,(\Output\,{\tt{false}},\tau) & \text{if } k_{in}\,p + k\,p \leq n < k_{in}\, + k\, p + k_{out}\,p
            \\&{\text{and}~}
            \tau = ({\tt{outputs\,p}})_n\\
            \Cell\, (\Internal, \tau) & \text{if } k_{in}\,p\leq n < k_{in}\,p + k\,p\\
            & {\text{and}~} 
            (v:\tau) = ({\tt{internals\,p}})_n,\\
            {\text{undefined}} & \text{otherwise}
        \end{cases}
    \end{array}
\end{equation*}
\begin{definition}
    A program \(p\) is well-typed if 
    \(\normalfont \aeval\,({\tt{program}}\,p)\,(\ainit\,p)\) is defined and
    the following conditions hold:
        \begin{itemize}
        \item for all \(n < k_{in}\) we have \(\normalfont (\aeval\,p\,(\ainit\,p))\, n = \Cell\,(\Input\,{\tt{false}},\cdot)\)
        \item for all \(k_{in} + k_{p} \leq n < k_{in} + k_{p} + k_{out}\) we have \(\normalfont (\aeval\,p\,(\ainit\,p))\, n = \Cell\,(\Output\,{\tt{false}},\cdot)\)
        \end{itemize}
\end{definition}
We conclude this section with the following results, which establish that well-typed programs
are reactive, i.e. they produce outputs for all possible inputs.
\begin{theorem}[Correctness]\label{theorem:correctness}
    Let \(p\) be a well-typed program and, \(i\) be a list compatible with \({\tt{inputs}}\,p\),
    and \(\sigma\) be a memory such
    that \(\normalfont {\pull}\,\,p\,\sigma\,i \in \gamma\,(\ainit\, p)\). Then, there exists \(\sigma'\) such that:
    \begin{itemize}
        \item \(\normalfont \eval\,({\tt{program}}\,p)\,{\tt{tt}}\,\sigma = ({\tt{tt}},\sigma')\)
        \item \(\normalfont \push\,\,p\,\sigma'\) is compatible with \({\tt{outputs}}\,p\)
        \item \(\normalfont {\pull}\,\,p\,\sigma'\,i' \in \gamma\,(\ainit\,p)\)
            for all \(i'\) compatible with \({\tt{inputs}}\,p\)
    \end{itemize}
\end{theorem} 
\begin{proof}
    First, prove that for all \(\sfval\), \(\sigma\), \(a\) and \(\Sigma\), 
    such that $\sigma \in \gamma\,\Sigma$ and $\aeval\, \sfval\,\Sigma = \Sigma'$ for
    some \(\Sigma'\), there exists \(\sigma'\) such that 
    \(\sfval\, a\,\sigma = (b,\sigma')\) and \(\sigma' \in \gamma\,\Sigma'\).
    The proof proceeds by induction on the structure of \(\sfval\).
    By \cref{lemma:res_preservation}, the abstract semantics preserves
    the classification of memory cells as internals, input or output.
    In particular, \(\sigma'\) remains compatible with \(\ainit\,p\) for all internal resources.
    Moreover, it is straightforward to verify that for all resource \(r\) of the program we have 
    \(r \vdash \ainit\,p\).
    From this, it follows that \({\pull}\,\sigma'\,'i \in \gamma\,(\ainit\,p)\)
    for all \(i'\) compatible with \({\tt{inputs}}\,p\). The final result follows by applying these properties
    to the body of \(p\).
\end{proof}
\begin{corollary}[Reactivity]
    Let \(p\) be a well-typed program. Then for all input streams 
    \(is : \StreamType\,({\tt{inputs}}\,p)\),
    there exists an output stream \(os : \StreamType\,({\tt{outputs}}\,p)\) such
    that \(\normalfont {\tt{run}}\,p\,(\init\,p)\,is = os\).
\end{corollary}
\begin{proof}
    By bisimulation on streams, using \cref{theorem:correctness}.
\end{proof}
For the remainder of this paper, we consider only well-typed programs applied to inputs of the correct type.
Consequently, based on results of this section, we can assume that the stepwise semantics and, by extension 
the semantics over streams, are total functions.

%% file: pages/equivalence.tex
In this section we state and prove the main result of this paper: 
the existence of a normal form for \ourlanguage{} terms, where 
all resource accesses occur at the top level, and all internal resources are read and written 
at most once. This first result is established in \cref{sec:yampanormalform}.
Furthermore, we show that these normal forms can be transformed into semantically equivalent 
\Yampacore{} normal forms. This second result is established in 
\cref{sec:yampatransformation}. 
We conjecture that similar results hold for \Wormholes{} terms, but to the best of our knowledge,
this has never been stated nor proven. Thus, our work extends previous work
not only by relaxing constraints on resource usage but also by establishing the
existence of normal forms and proving their equivalence to \Yampacore{} terms.
Another result, not detailed here for brevity, is that \Yampacore{} normal forms 
can also be transformed into \ourlanguage{} normal forms.
Together, these results, all of which rely on the equational theories presented in previous sections and on bisimulation proofs,
establish that \ourlanguage{} remains compatible with existing functional
paradigms. 

\subsection{Normal form}
\label{sec:yampanormalform}

A term $t$ may contain a set of resources which can be separated between the read ones and the written ones. We define $\refget$, respectively $\refset$, a function that takes a term $t$ and returns an ordered list of resources used for read, respectively for write, without multiple occurrences. With these functions declared, we state that a term in normal form has the following structure
\[\Get\,r'_0 \compop \ldots \compop \Get\,r'_{n_g-1} \compop \Arr\,f \compop \Set\,r'_0 \compop \ldots \compop \Set\,r'_{n_s-1}\]
with function $f$ from the host language, and $n_g = |\refget(t)|$ and $n_s = |\refset(t)|$. For readability, we use the infix operator $(\compop)$ instead of $\Comp$, which is left associative.
Knowing that, resource identifiers for inputs are lower than ones for internals, and internals ones are lower than outputs ones, we can add that input and output operations are at the outermost of the term. 
\begin{example}
For instance, defining $\mathit{add} = \lambda(x,y). x + y$, we expect to prove that the following holds:
\[\Arr\,(\dup) \compop \Set\,r' \compop \Get\,r \compop \Arr\,\mathit{add} \equiv \Get\,r \compop \Arr\,f \compop \Set\,r'\]
where
\(f = \lambda (x,y). \letin{x'}{\dup\,x}{\letin{(w,z)}{\perm~(x',y)}{(\mathit{add}~w,z)}}\).
\end{example}
\begin{figure}[ht]
  \begin{align*}
    \stackfirst(n,t) &= \overbrace{\First\,(\ldots\,(\First}^{n}\,t)\ldots) \\
    \permlf(n,t) &= \stackfirst(n - 1,\Arr\,\perm) \compop \ldots \compop \stackfirst(0,\Arr\,\perm) \compop t \\
    \permrf(n,t) &= t \compop \stackfirst(0),\Arr\,\perm) \compop \ldots \compop \stackfirst(n-1,\Arr\,\perm) \\
    \complg(l,t) &= (\Get\,r_0 \compop (\ldots \compop (\Get\,r_{|l|-1} \compop t) \ldots )) \\
    \comprg(l,t) &= t \compop \Get\,r_{|l|-1} \compop \ldots \compop \Get\,r_{0} \\
    \complg(l,t) &= (\Set\,r_0 \compop (\ldots \compop (\Set\,r_{|l|-1} \compop t) \ldots )) \\
    \comprs(l,t) &= t \compop \Set\,r_{|l|-1} \compop \ldots \compop \Set\,r_{0} 
  \end{align*}
  \vspace{-0.5cm}
  \caption{Concise notation for parameterized terms}
  \label{fig:concise:notation}
\end{figure}
The rest of this section is dedicated to proving the normal form theorem. 
Before proceeding with the proof, we introduce several intermediary results 
and define notations that enable a concise representation of terms involving sequences of $\First$ constructs, resources accesses, and permutation arrows. These notations are detailed in \cref{fig:concise:notation}. We define $\mathit{concat}$ as the function that takes two lists and returns a new list containing all the resources from both, while maintaining increasing order and ensuring
that each resources appears only once.

A term is said to effect-free if it does not contain any $\Get$ or $\Set$ operation.
\cref{lemma:contraction} states that any effect-free term can be contracted to a simple arrow. \cref{lemma:jump_over} shows that a sequence of read, or write, accesses can be moved over a $\First$ constructor. \cref{lemma:swap} states that we can swap
sequences of read and write operations, while \cref{lemma:merge} states that we can merge two sequences of read operations or two sequences of write operations.
For brevey, we omit the proofs of these lemmas. In each case, the proof proceeds by
structural induction on lists, except for \cref{lemma:contraction} which is proven by structural induction on the term.
\begin{lemma}[Contraction]
  \label{lemma:contraction}
  For any effect-free term $t$, there exists a function $f$ in the host language such that $t \equiv \Arr\,f$.
\end{lemma}
\begin{lemma}[Jump over]
  \label{lemma:jump_over}
  For any term $t$ and any list of resources $l$,
  \begin{align*}
      \First\,(\complg(l,t)) &\equiv \complg(l,\permlf(|l|,\First\,t)) \\
      \First\,(\comprs(l,t)) &\equiv \comprs(l,\permrf(|l|,\First\,t))
  \end{align*}
\end{lemma}
\begin{lemma}[Swap]
  \label{lemma:swap}
  For any function $g$ and any two list of resources $l$ and $l'$, there exists a function $f$ in the host language such that
  \[\comprg(l_1,\compls(l_2,\Arr\,g)) \equiv \comprs(l'_2,\complg(l'_1,\Arr\,f))\]
  where $\forall r \in l'_2, r \in l_2 \wedge r \notin l_1$ and $\forall r \in l'_1, r \in l_1 \wedge r \notin l_2$. 
  Additionally, if $l_1$ and $l_2$ are ordered and contain no duplicate resources, then $l'_1$ and $l'_2$ also retain this property.
\end{lemma}
\begin{lemma}[Merge]
  \label{lemma:merge}
  For any function $g$ and two list of resources $l$ and $l'$, there exists a function $f$ such that
  \begin{align*}
      \complg(l,\complg(l',\Arr\,g)) &\equiv \complg(\mathit{concat}(l,l'),\Arr\,f) \\
      \comprs(l,\comprs(l',\Arr\,g)) &\equiv \comprs(\mathit{concat}(l,l'),\Arr\,f)
  \end{align*}
\end{lemma}
We conclude this section by stating and proving the normal form theorem.
\begin{theorem}[Normal form]
    For any term $t$, there exists a function $f$ of the host language such that 
    \text{$t \equiv \comprs(\refset(t),\complg(\refget(t),\Arr\,f))$}, i.e.
    \[t \equiv \Get\,r_0 \compop \ldots \compop \Get\,r_{n_g-1} \compop \Arr\,f \compop \Set\,r'_0 \compop \ldots \compop \Set\,r'_{n_s-1}\]
    where $n_g = |\refget(t)|$ and $n_s = |\refset(t)|$.
\end{theorem}
\begin{proof}
    The proof is by structural induction on the term $t$.
    \begin{itemize}
        \item $\Arr\,f \equiv \Arr\,f$ by reflexivity
        \item $\Get\,r \equiv \complg([r],\Arr\,id) = \Comp\,(\Get\,r)\,(\Arr\,id)$ \text{(\cref{eq:arrowlaw2})}
        \item $\Set\,r \equiv \complg([r],\Arr\,id) = \Comp\,(\Arr\,id)\,(\Set\,r)$ \text{(\cref{eq:arrowlaw1})}
        \item If $t \equiv \comprs(\refset(t),\complg(\refget(t),\Arr\,f))$ then
        \begin{align*}
            \First\,t \equiv~& \First\,\comprs(l^{\tt S},\complg(l^{\tt G},\Arr\,f)) &(\text{\cref{lemma:equiv:Molholes:congruence}})\\
            \equiv~& \comprs(l^{\tt S},\permrf(s^{\tt S},\First\,(\complg(l^{\tt G},\Arr\,f)))) &(\text{\cref{lemma:jump_over}}) \\
            \equiv~& \comprs(l^{\tt S},\permrf(s^{\tt S},\complg(l^{\tt G},\permlf(s^{\tt G},\First\,(\Arr\,f))))) &(\text{\cref{lemma:jump_over}}) \\
            \equiv~& \comprs(l^{\tt S},\complg(l^{\tt G},\permrf(s^{\tt S},\permlf(s^{\tt G},\First\,(\Arr\,f))))) &(\text{\cref{eq:arrowlaw3}})
        \end{align*}
        where $l^{\tt S} = \refset(t)$, $l^{\tt G} = \refget(t)$, $s^{\tt S} = |l^{\tt S}|$ and $s^{\tt G} = |l^{\tt G}|$. By \cref{lemma:contraction}, there exists $g$ such that $\Arr\,g \equiv \permlf(s_s,\permlf(s_g,\First\,(\Arr\,f)))$. Consequently, the equivalence $\First\,t \equiv \comprs(l_s,\complg(l_g,\Arr\,g))$ holds.
        
        \item If $t_1 \equiv \comprs(l^{\tt S}_1,\complg(l^{\tt G}_1,\Arr\,f_1))$ and $t_2 \equiv \comprs(l^{\tt S}_2,\complg(l^{\tt G}_2,\Arr\,f_2))$ then by \cref{lemma:equiv:Molholes:congruence} we can rewrite $\Comp\,t_1\,t_2$ as
        \[\Comp\,(\comprs(l^{\tt S}_1,\complg(l^{\tt G}_1,\Arr\,f_1)))\,(\comprs(l^{\tt S}_2,\complg(l^{\tt G}_2,\Arr\,f_2)))\]
        where $l^{\tt G}_1 = \refget(t_1)$, $l^{\tt G}_2 = \refget(t_2)$, $l^{\tt S}_1 = \refset(t_1)$ and $l^{\tt S}_2 = \refset(t_2)$. This case relies on straightforward transformations
        that allow us to focus on specific sub-terms, enabling the application of \cref{lemma:merge,lemma:swap}. However, due to space constraints, we do explicitly state these transformations. They can be justified by multiple applications of \cref{eq:arrowlaw1,eq:arrowlaw2,eq:arrowlaw3}. First, we focus on the sequence of $\Set$ operations in $t_1$ and the sequence of $\Get$ operations in $t_2$. This restructuring leads to the following term:
        \[\Comp\,(\Comp\,(\complg(l^{\tt G}_1,\Arr\,f_1))\,(\comprg(l^{\tt G}_2,\compls(l^{\tt S}_1,\Arr\,id))))\,(\comprs(l^{\tt S}_2,\Arr\,f_2))\]
        Now we apply \cref{lemma:swap}, which allow us to swap sequences by introducing
        a term \(\Arr\,f\) between them. During this process, some resources accesses may
        be eliminated.
        Thus, for some lists \(l_i'^{\tt S}\) which may be subsets of \(l_i^{\tt{S}}\), we obtain the following term:
        \[\Comp\,(\Comp\,(\complg(l^{\tt G}_1,\Arr\,f_1))\,(\comprs(l'^{\tt S}_1,\complg(l'^{\tt G}_2,\Arr\,f))))\,(\comprs(l^{\tt S}_2,\Arr\,f_2))\]
        By a straightforward transformation,
        we decompose this term into two sub-terms, ensuring that all $\Get$ appear on the left and all $\Set$ appear on the right which leads to the equivalent term
        \[\Comp\,[\Comp\,(\complg(l^{\tt G}_1,\Arr\,f_1))\,(\complg(l'^{\tt G}_2,\Arr\,f))]\,
                 [\Comp\,(\comprs(l'^{\tt S}_1,\Arr\,id))\,(\comprs(l^{\tt S}_2,\Arr\,f_2))]\]
        The left-hand side of the top-level composition can be rewritten as 
        \[(1)\,\complg(l^{\tt G}_1,(\complg(l'^{\tt G}_2,\Comp\,\stackfirst(|l'^{\tt G}_2|,\Arr\,f_1)\,(\Arr\,f))))\]
        By \cref{lemma:contraction}, there exists a function $g$ such that $\Arr\,g \equiv \Comp\,\stackfirst(|l'^{\tt G}_2|,\Arr\,f_1)\,(\Arr\,f)$. Furthermore, by \cref{lemma:merge} there exists a function $g_1$ such that \(1\) can be rewritten as $\complg(\mathit{concat}(l^{\tt G}_1,l'^{\tt G}_2),\Arr\,g_1)$.
        By following the same approach for the right-hand side of the top-level composition, there exists
        a function \(g_2\) such that the entire term can be rewritten as
        \[\Comp\,(\complg(\mathit{concat}(l^{\tt G}_1,l'^{\tt G}_2),\Arr\,g_1))\,(\comprs(\mathit{concat}(l'^{\tt S}_1,l^{\tt S}_2),\Arr\,g_2))\]
        This term simplifies further to
        \[\comprs(\mathit{concat}(l'^{\tt S}_1,l^{\tt S}_2),\complg(\mathit{concat}(l^{\tt G}_1,l'^{\tt G}_2),\Comp\,\Arr\,g_1\,\Arr\,g_2))\]
        which, by applying \cref{eq:arrowlaw4}, leads to the expected normal form.
    \end{itemize}
\end{proof}
\def\translate{{\mathit{translate}}}
\subsection{Transformation into \Yampa{}}
\label{sec:yampatransformation}
From a program \(p\) obtained through the previous transformation, it is  possible to construct an equivalent version where all
resources of the same kind are merged into a single resource, and all read (resp. write) accesses 
to resources of the same kind are merged into a single read (resp. write) access. 
For lack of space, we do not detail this transformation here. Instead, we assume that the
program obtained from the previous section takes the following form:
\[ 
    \begin{array}{lll}
        p' =&   \{ & {\tt{inputs}} = [((\tau_1 \times \tau_2) \times \ldots) \times \tau_n];\qquad
           {\tt{outputs}} = [((\tau'_{1} \times \tau'_{2}) \times \ldots) \times \tau'_{l}];\\
           && {\tt{internals}} = [(((v_{n+1}:\tau_{n+1}, v_{n+2}:\tau_{n+2}), \ldots), v_{n+m}:\tau_{n+m})];\\
           && {\tt{program}} = 
           \Comp\, (\Get\,r_{in})\,(\Comp\,(\Get\,r)\,(\Comp\,(\Arr\,f)\,(\Comp\,(\Set\,r)\,(\Set\,r_{out})))) \}
    \end{array}
\]
where \(n = k_{in}\,p\), \(m = k\,p\) and \(l = k_{out}\,p\).
Types and values results from collapsing their counterpart in \(p\)
and, \(r_{in}\),\(r\) and \(r_{out}\) are the single input, internal and output resources, respectively.
The function \(f\) has type 
\(((\VoidType \times {\tt{inputs}}\,p') \times {\tt{internals}}\,p') \rightarrow
((\VoidType \times {\tt{outputs}}\,p') \times {\tt{internals}}\,p')\).

Given a program \(p\) of the form described above, 
we define 
\[
C_p = \{\sigma \mid {\text{if \(i\) is compatible with \({\tt{inputs}}\,p\) then }}
\sigma \in \gamma\,(\ainit\,p)\}
\]
Up to permutation of the parameters of 
\(\eval\,({\tt{program}}\,p)\), the pair \((C_p\, \eval\,({\tt{program}}\,p))\) forms a 
coalgebra for the functor \(F_{IO}\), where \([I] = {\tt{inputs}}\,p\) and \([O] ={\tt{outputs}}\,p\).
Thus, we can define a bisimulation between \((C_p,\eval\,({\tt{program}}\,p))\) 
and \((sf\,I\,O,\,{\mathit{id}})\). A relation \(R\) is a bisimulation if for all
\(\sigma \in C_p\), \(a : I\), \(b : O\) and \(\sfval : \sfType\,I\,O\)
such that \(R\,\sigma\,sf\),
\[
\begin{array}{l}
    {\text{if }} \eval\,({\tt{program}}\,p)\,\void\,({\pull}\,p\,\sigma\,[a])= (\void,\sigma') 
    {\text{ then there exists \(\sfval'\) such that }}\\
    \qquad \sfval\,a= (b,\sfval'),
    {\text{where }} [b]={\push}\,p\,\sigma', \text{ and } R\,\sigma'\,\sfval'
\end{array}
\]
The transformation of a program \(p\) into \Yampa{} is expressed as 
\({\mathit{translate}}\,p\,({\tt{init}}\,p)\)
 where, \(f\) being  the function appearing in the definition of \(p\), and \(v\) the value
that aggregates the initial values of the internal resources, the function \(\translate\) is defined as:
\[
    {\mathit{translate}}\,p\,\sigma = \Loop\,v\,(\Arr\,f)
\]
With this definition, the initial memory of the program \(p\) is bisimilar to the
semantics of its translation (\cref{lemma:finaleq}). As a corollary, we conclude that \(p\) and its
transformation denote the same stream function (\cref{corollary:finaleq}). 
\begin{proposition}
    \label{lemma:finaleq}
    For all well-typed program \(p\), the following holds: 
    \[\normalfont \init\,p \sim
    \eval\,(\translate\,p\,(\init\,p))\]
    where \(\sim\) denotes bisimilarity between \(\normalfont (C_p,\eval\,({\tt{program}}\,p))\) and 
    \((\sfType\,I\,O,{\mathit{id}})\).
\end{proposition}
\begin{proof}
    First, prove that if
    \(\eval\,(\translate\,p\,\sigma)\,a = (b,\sfval)\) then 
    \(\sfval = \translate\,p\,\sigma'\) for some appropriate \(\sigma'\). 
    The key observation is that both \(\lambda a.\eval\,({\tt{program}}\,p)\,\void\, 
    ({\mathit{pull}}\,p\,\sigma\,a)\) 
    and \(\translate\,p\,\sigma\) 
    apply the same function \(f\) to values
    occurring in the inputs on one side, and values of internal resources and loop states on
    the other side.
    Thus, the resulting values are the same of both side, up to a shift between
    inputs on one side and internal resources and loop states on the other side.
    Next, prove that \(R=\{(\sigma,\eval\,(\translate\,p\,\sigma)) \mid \sigma \in C_p\}\)
    is a bisimulation.
\end{proof}
\begin{corollary}
    \label{corollary:finaleq}
    For any well-typed program \(p\), \(p\) and \(\normalfont \translate\,p\,(\init\,p)\) 
    denote the same stream function, i.e.
    \(\normalfont
        \run\,p\,(\init\,p)\,s \sim \run\,(\translate\,p\,(\init\,p))\,s
    \)
    where \(\sim\) denotes bisimilarity between streams.
\end{corollary}
\begin{proof}
    This follows by
    applying \cref{lemma:finaleq}, along with the definitions of the two {\tt{run}} functions 
    and bisimulation, and
    the bisimulation principle for streams.
\end{proof}
As mentioned earlier, one can also show that the transformation between normal forms
is reversible, thereby establishing that \ourlanguage{} and \Yampacore{}
are equivalent in terms of expressivity.

%% file: pages/conclusion.tex
We have introduced \ourlanguage, a programming language that extends \Wormholes by relaxing constraints on resource usage. Specifically, \ourlanguage permits multiple reads and writes to internal resources while still enforcing the single-use policy for input and output resources. A type system ensures correct resource usage.

We have demonstrated that well-typed programs can be transformed into equivalent versions where all resources are accessed at most once and at the top level. These transformations are justified using the equational theories of categories, functors, monads, and arrows, along with additional equalities specific to the state monad used in the language’s semantics.

Building on this transformation, we have shown that \ourlanguage programs can be translated into equivalent resource-free \Yampacore programs. Since the semantics of \Yampacore arrows are defined using corecursive functions, this equivalence is established via bisimulation proof techniques. Together, these results confirm that our approach preserves the functional nature of \Yampa while offering a more flexible resource management system than \Wormholes.

This work was initially motivated by the development of a reactive programming library (currently in progress) for the \OCaml language~\cite{OCAML:MISC}, integrating \Yampa's programming model with \OCaml's impure nature.

For future work, we plan to formalize our results using the \Coq proof assistant~\cite{COQ:MISC} and extract certified programs from the formalization. Our long-term goal is to verify program correctness against behavioral specifications, potentially expressed in temporal logic, enabling the generation of certified reactive programs. We expect the
semantics domain of \ourlanguage to be expressive enough to define the semantics of similar languages,
i.e., languages that combine the dataflow approach with effectful computations.

Another research direction involves extending \ourlanguage with additional constructs. The combinators discussed in this paper cover only a restricted subset of signal functions, which simplifies bisimulation due to the limited update behavior in computation steps. We aim to investigate \Yampa combinators beyond this restriction and analyze their impact on bisimulation proofs. Additionally, we plan to explore language extensions that enhance expressivity, including other computational effects, such as exception handling, to improve compatibility with \OCaml. In particular, the interaction between resources and \OCaml's references is of interest.